\documentclass[12pt,letterpaper]{article}
\usepackage[top=2.5cm, bottom=2.5cm, left=2.5cm, right=2.5cm]{geometry}
\usepackage[utf8]{inputenc}
\usepackage{setspace}
\usepackage[T1]{fontenc}
\usepackage{lmodern} %  Latin modern fonts in outline formats
\usepackage{graphicx,subfigure}
\usepackage{siunitx}
\usepackage{booktabs}
\usepackage{textcomp} % For additional text symbols and Unicode support
\usepackage{float,fancybox,adjustbox,colortbl}
\usepackage{amssymb} % For math environment bold format
\usepackage{lineno} % line numbers
\usepackage{multirow,multicol}

\usepackage{color}
\usepackage[table,xcdraw]{xcolor}
\usepackage{bbm}
\usepackage{amsmath,bm,amsthm}% Maths environments
\usepackage{subcaption}
\usepackage{authblk} % Author title
\usepackage{nameref}
\usepackage{fancyhdr}
\usepackage{pdflscape}
\usepackage{url}
\usepackage{gensymb} % Generic symbols for both text and math mode

\usepackage{authblk} % Author title
\usepackage{nameref}
\usepackage{fancyhdr}

\newcommand{\beginsupplement}[1]{%
        \setcounter{table}{0}
        \renewcommand{\thetable}{#1.\arabic{table}}%
        \setcounter{figure}{0}
        \renewcommand{\thefigure}{#1.\arabic{figure}}%
        \setcounter{section}{0}
        \setcounter{subsection}{0}

        \setcounter{section}{0} % Resetting section counter
        \setcounter{subsection}{0} % Resetting subsection counter
        \renewcommand{\thesection}{S\arabic{section}} % Prefix with 'S' for supplementary
        }
 \usepackage[vlined,ruled]{algorithm2e}
\definecolor{darkblue}{rgb}{0.0, 0.0, 0.65}
\usepackage[pdftex,colorlinks=true,linkcolor=darkblue,citecolor=darkblue,urlcolor=blue]{hyperref}
\usepackage[square,numbers]{natbib}%[authoryear,sort]
\newcommand*{\doi}[1]{\href{http://dx.doi.org/#1}{doi: #1}} % Add hiperlink to doi

\usepackage[toc,page]{appendix}

\makeatletter
\renewcommand{\maketitle}{\bgroup\setlength{\parindent}{0pt}
\begin{flushleft}
  \textbf{\@title}\\ \vspace{0.5cm}
  \textbf{\@author}
\end{flushleft}\egroup
}
\makeatother

\usepackage{soul} % for strike out: replace
\definecolor{Gray}{gray}{0.8}

\font\myfont=cmr12 at 14pt

\title{ \myfont Assessing Methodological Variability in Wastewater Surveillance: A Wavelet Decomposition Approach}
\author[1, +]{Maria L. Daza–Torres}
\author[1, +]{J. Cricelio Montesinos-L\'opez}
\author[3]{Rachel Olson}
\author[3]{ C. Winston Bess }
\author[2]{Colleen C. Naughton}
\author[3,*]{Heather N. Bischel}
\author[1,*]{Miriam Nu\~no}

\affil[1]{Department of Public Health Sciences, University of California Davis, Davis, California, USA}
\affil[2]{Department of Civil and Environmental Engineering, University of California Merced, Merced, California, USA}
\affil[3]{Department of Civil and Environmental Engineering, University of California Davis, Davis, California, USA}
\affil[4]{Department of Internal Medicine, University of California Davis, Sacramento, California, USA}
\affil[+]{co-first author}
\affil[*]{co-corresponding authors: mnuno@health.ucdavis.edu; hbischel@ucdavis.edu}
\begin{document}

\maketitle
%\linenumbers
\begin{abstract}
Wastewater surveillance has emerged as a critical public health tool, enabling early detection of infectious disease outbreaks and providing timely, population-level insights into community health trends. However, variability in sample collection and processing, for example between wastewater influent and settled solids, can introduce methodological noise that differentially impacts true epidemiological signals and limits cross-site comparability.

To address this challenge, we aimed to discern underlying disease trends from methodological variability in SARS-CoV-2 wastewater data using discrete wavelet transform (DWT), with a focus on comparing influent and solids samples from the same geographic locations.

We applied DWT to longitudinal SARS-CoV-2 RNA concentrations in wastewater from five cities in California: Los Banos, Turlock, Woodland, Winters, and Esparto—each with paired influent and solids samples. DWT decomposes each signal into two components: (1) approximation coefficients that capture smoothed long-term trends, and (2) detail coefficients that isolate high-frequency fluctuations and transient variations in the signal. We reconstructed signals by progressively removing the high-frequency components (detail coefficients) and assessed similarity between sample types using hierarchical clustering.

Clustering of raw signals did not yield city-specific groupings, indicating that methodological noise obscured the underlying epidemiological signal. Intermediate reconstructions that retained some high-frequency components continued to show mixed groupings. In contrast, reconstructions based solely on low-frequency approximation coefficients revealed clear, city-specific clustering, with influent and solids samples from the same city aligning closely.

These findings support our hypothesis that high-frequency components are primarily driven by sample processing and laboratory noise, while low-frequency components reflect shared epidemiological trends. Our findings underscore the importance of denoising in wastewater data preprocessing and offer a scalable approach for enhancing signal comparability across regions and sample types.

\end{abstract}

\section{Introduction}
Wastewater-based epidemiology (WBE) is a valuable tool for monitoring severe acute respiratory syndrome coronavirus 2 (SARS-CoV-2) shedding within populations. It provides a cost-effective, scalable, and non-invasive method for tracking disease transmission trends \cite{kumblathan2021wastewater, hart2020computational}. Numerous studies have demonstrated the utility of WBE in complementing clinical data, providing early warning signals of outbreaks, and enabling real-time surveillance during the Coronavirus disease 2019 (COVID-19) pandemic \cite{peccia2020measurement,shah2022systematic,DAZATORRES2024112485}. 

WBE still lacks universally adopted protocols for sampling, virus concentration, and detection. Methodologies vary substantially across laboratories, including differences in sampling methods (e.g., influent vs. settled solids), virus concentration methods, RNA extraction techniques, and detection assays such as ddPCR or next-generation sequencing \cite{mcclary2021standardizing,mcclary2021sars, cuadros2024advancing}. This variability introduces substantial noise and uncertainty into the data, complicating efforts to compare results across sites and limiting the potential for regional or national-scale integration \cite{cuadros2024advancing, mcclary2021sars}. As a result, true temporal trends may be obscured, and the timely detection of emerging outbreaks can be delayed.  Although standardization efforts are underway \cite{cuadros2024advancing},  the influence of specific technical factors on signal interpretation remains inadequately quantified, posing ongoing challenges for harmonizing wastewater surveillance data.
Effectively isolating the underlying epidemiological signal from this methodological noise requires techniques capable of decomposing the data across temporal scales, distinguishing long-term trends from transient short-term fluctuations.

The discrete wavelet transform (DWT) is especially well-suited for analyzing non-stationary time series typical of environmental and epidemiological data \cite{debnath2015wavelet, daubechies1992ten}. DWT decomposes signals into components localized in both time and frequency, allowing us to separate smooth, low-frequency trends (approximation coefficients) from short-term, high-frequency fluctuations (detail coefficients) that may arise from laboratory or sampling variability. Wavelet-based analysis has gained significant popularity in analyzing time series data, with numerous applications across various domains, including climate analysis, ecology, epidemiology, biology, and water quality monitoring  \cite{garcia2023common, cazelles2014wavelet, deniz2025analysis, guo2022review}. In infectious disease modeling, wavelets have been used to detect periodic patterns, temporal shifts in transmission, and early signs of outbreaks \cite{cazelles2014wavelet, grenfell2001travelling, Minaeian2023}. In water quality analysis,  wavelet models have demonstrated their ability to detect trends and anomalies in complex datasets, including indicators like salinity, nitrate concentration, chemical oxygen demand, and total suspended solids \cite{w12051476}.

This study uses wavelet analysis to address signal variability resulting from differences in sample matrix, specifically between influent and settled solids. These matrices differ in their physical and chemical properties, viral RNA stability, and viral load, all of which can affect signal strength and interpretation \cite{graham2020sars}. We hypothesize that sample type variability manifests predominantly in the high-frequency components of the signal. Thus, removing these components via wavelet denoising should reveal a shared underlying epidemiological trend.

To test this hypothesis, we analyze longitudinal SARS-CoV-2 wastewater data from five cities in California, Los Banos, Turlock, Woodland, Winters, and Esparto, for which we collected and analyzed both influent and settled solids samples. Our approach includes three steps. First, we reconstruct progressively smoothed versions of each signal by systematically removing high-frequency wavelet coefficients from the highest to the lowest frequency bands. Second, we compare clustering patterns of original versus denoised signals to assess how methodological noise influences observed similarity. Lastly, we evaluate whether denoised signals exhibit stronger clustering by geographic location than by sample type.

This approach provides a novel framework for disentangling community-level infection dynamics from sample-type variability, thereby enhancing the interpretability and the potential for integrating wastewater surveillance data across sites with different methods.

%%%%%%%%%%%%%%%%%%%%%%%%%%%%%%%%%%%%%%%%%%%%%%%%%%%%%%%%%%%%%%%%%%%%%%
%%%%%%%%%%%%%%%%%%%%%%%%%%%%%%%%%%%%%%%%%%%%%%%%%%%%%%%%%%%%%%%%%%%%%%
\section{Sample Collection and Processing}
We analyzed SARS-CoV-2 wastewater data from influent and settled solids samples collected as part of the Healthy Central Valley Together (HCVT) project \cite{Kadonsky2023} between April 21 and November 30, 2022 (224 days). Five wastewater treatment plants (WWTPs) in Central Valley cities in California (Los Banos, Turlock, Woodland, Winters, and Esparto) were included in the present study. Twenty-four hour composite raw wastewater samples (~1L in Nalgene HDPE bottles) were collected four to five days per week for each facility over the study period. Each raw wastewater sample was then processed via two pipelines (influent and settled solids from the influent). The timeframe corresponds to the period of maximal temporal overlap for both sample types across all five locations. Table \ref{tab:data_descrip} summarizes key characteristics for each city included in the study, detailing the estimated population served, wastewater capacity, and the total number of influent and solids samples collected over the study period.

A 50-mL aliquot of each raw wastewater sample was taken for the influent processing pipeline. The methodology for processing and analyzing influent samples is described in  \cite{DAZATORRES2022159680} and \cite{muralidharan2025equity}. In brief, the influent processing protocol included pasteurization for 30 minutes at 60 °C, spiking the sample with bovine coronavirus (BCOV) as a recovery control, concentration of viruses from 4.875-ml sample aliquots using Nanotrap® Magnetic Virus Particles (Ceres Nanosciences, Inc.), magnetic bead-based RNA and DNA extraction, and analysis of virus targets by reverse transcription droplet digital polymerase chain reaction (RT-ddPCR). The virus targets included SARS-CoV-2 RNA (N-gene) and pepper mild mottle virus (PMMoV). 

Settled solids were obtained from each raw wastewater sample by allowing the remaining solid material to settle in a beaker for at least 30 minutes and decanting the overlying water. Settled solids were then transferred to centrifuge tubes and dewatered by centrifugation at 24,000×g for 30 min at $90^\circ$C. Dewatered solids samples were analyzed as detailed in \cite{Kadonsky2023}. The protocols for sample preparation, RNA extraction, and RT-ddPCR were based on methods developed and applied by \cite{loeb2020extraction,wolfe2022detection,wolfe2021,Topol2022v5,topol2021high}. In brief, the dewatered solids processing protocol included spiking the sample with BCOV as a recovery control, magnetic bead-based extraction of RNA and DNA, and PCR inhibitor removal.

\begin{figure}[H]
\begin{center}
\includegraphics[scale=0.42]{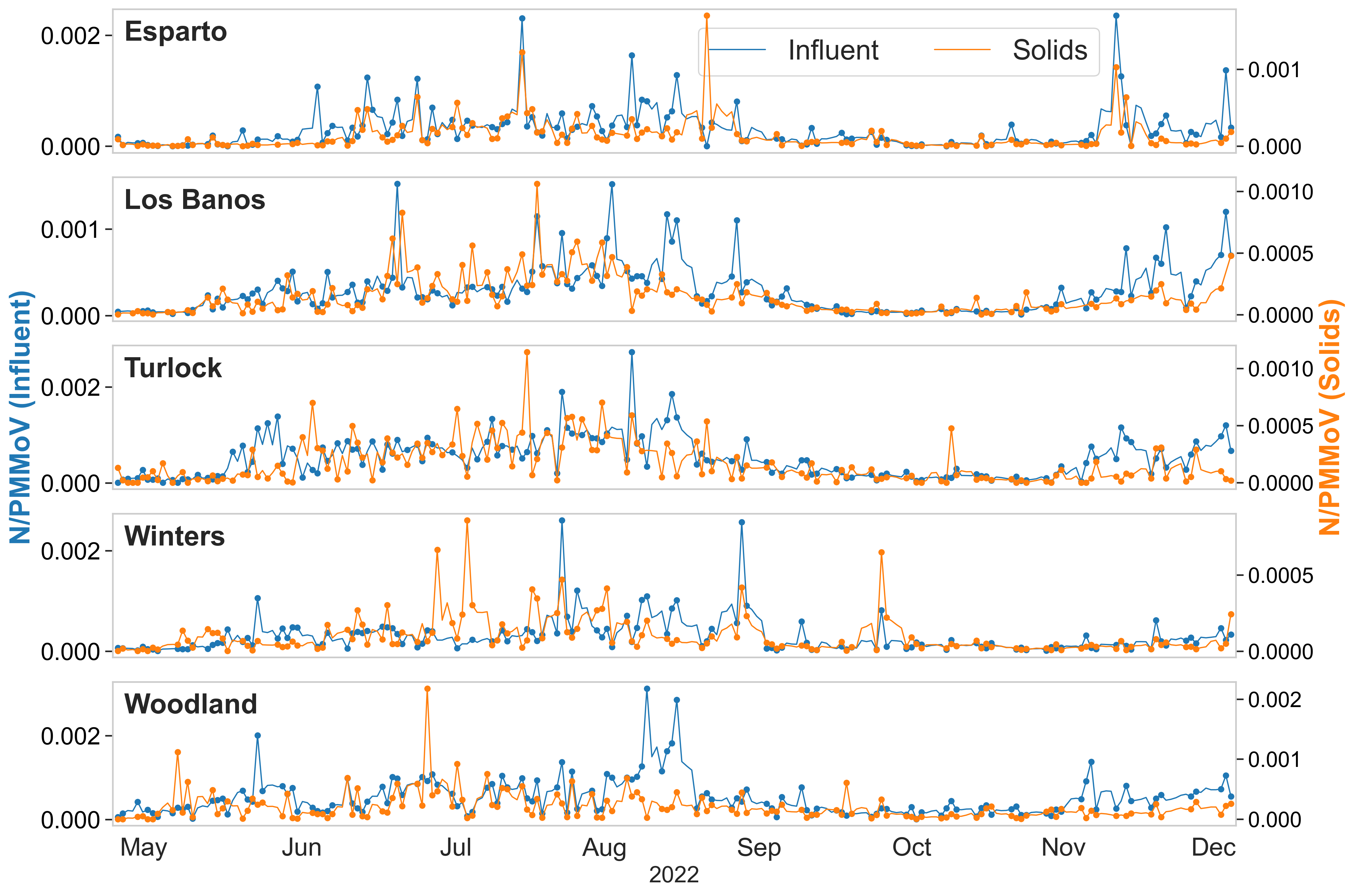}
    \caption{Time series of normalized SARS-CoV-2 concentrations in wastewater influent and settled solids samples from Esparto, Los Banos, Turlock, Winters, and Woodland. Dots represent raw data. Time points without data collection were imputed using a 10-day moving average, resulting in the continuous lines that interpolate between gaps.}
    \label{fig:influent_vs_solids_ts}
\end{center}
\end{figure}

%\subsection{Data Normalization and Processing}
To help account for population dynamics and wastewater flow variations, SARS-CoV-2 RNA concentrations were normalized using PMMoV, a fecal indicator and process control. Normalization produces a dimensionless metric, represented as N/PMMoV, which we refer to as the normalized wastewater concentration or simply wastewater signal. Prior mass-balance modeling studies suggest that this ratio is proportional to the number of individuals shedding SARS-CoV-2 RNA in the contributing population \citep{doi:10.1128/msystems.00829-21}. Figure \ref{fig:influent_vs_solids_ts} displays normalized wastewater concentrations across all cities. Time points without data collection were imputed using a 10-day moving average, resulting in the continuous lines that interpolate between gaps.

\begin{table}[H]

\caption{Operational characteristics of WWTPs and sample coverage. Estimated population served, capacity (in million gallons per day), and the number of SARS-CoV-2 measurements collected from influent and solids samples. }
\label{tab:data_descrip}
\centering
\resizebox{0.8\textwidth}{!}{
\begin{tabular}{l l r r r r}
\toprule
\textbf{City} & \textbf{County} & \textbf{Population} & \textbf{Capacity (MGD)} & \textbf{Influent (N)} & \textbf{Solids (N)} \\
\midrule
Esparto & Yolo & 3,272 & 0.36 & 130 & 132 \\
Los Banos & Merced & 42,000 & 4.8 & 136 & 136 \\
Turlock & Stanislaus & 86,000 & 20 & 138 & 142 \\
Winters & Yolo & 7,200 & 0.92 & 125 & 129 \\
Woodland & Yolo & 59,000 & 10.4 & 136 & 138 \\
\bottomrule
\end{tabular}}
\end{table}

\section{Data Processing Methods}
We briefly present the theoretical foundation of wavelet transforms and outline our implementation of discrete wavelet analysis for comparing signal structures across sample types (See \cite{debnath2015wavelet}, for a full theoretical overview).

\subsection{Wavelet Transform}
A wavelet transform is the decomposition of a signal into a set of basis functions consisting of contractions, expansions, and translations of a mother function $\psi(t)$, called the wavelet.
The continuous wavelet transform (CWT) of a signal $x(t)$ is defined as:

$$W_x (s,\tau)= \frac{1}{\sqrt s}\int_{-\infty}^\infty x(t)\psi^*_{s,\tau}(t)dt,$$
where $$\psi_{s,\tau}(t)=\frac{1}{\sqrt s} \psi\left(\frac{t-\tau}{s}\right)$$
is a scaled and translated version of  $\psi(t)$, and  $\psi^*$ denotes its complex conjugate. The scale parameter $s$ controls frequency resolution, while the translation parameter $\tau$ controls temporal localization.
% ($s=2^j$, $\tau=k\cdot 2^j$)
Although the CWT offers a fine-grained and redundant representation, it is computationally intensive. The discrete wavelet transform (DWT) addresses this by restricting the scale and translation parameters to dyadic values, yielding an orthogonal, non-redundant representation that is computationally efficient and particularly effective for denoising and structure preservation \cite{WHITE20011419}.

\subsubsection{Discrete Wavelet Decomposition}
The DWT represents a discrete time series $\mathbf{x} = [x_1, \ldots, x_N]$ as a set of wavelet coefficients sampled from the CWT. These coefficients are computed using a filter bank consisting of a pair of finite impulse response filters: a low-pass filter $g$, which captures coarse approximation features, and a high-pass filter $h$, which extracts fine-scale details.

At each decomposition level $l$, the signal is processed recursively. Convolution with the low-pass filter yields the approximation coefficients:
$$cA_l[n] = \sum_{k=1}^{N/2^l} cA_{l-1}[k] \cdot g[2n - k], $$
while convolution with the high-pass filter produces the detail coefficients:
$$\; cD_l[n] = \sum_{k=1}^{N/2^l} cA_{l-1}[k] \cdot h[2t - k],$$
where $cA_0= \mathbf{x} $, the original signal. This recursive process builds a hierarchical, multiscale representation of the signal. The approximation coefficients reflect the smoothed, long-term components of the signal, while the detail coefficients capture high-frequency, transient variations at level $l$. 
% The length of coefficients at each level follows the dyadic scaling: $N / 2^l$, where $N$ is the length of the original series and $l\in \{1,...,l\}$.
\subsection{Wavelet Implementation}

%\subsubsection{Basis Selection}
The multiscale decomposition in the DWT framework depends on the choice of wavelet basis, determined by the filters $g$ and $h$, which define the underlying scaling and wavelet functions, respectively. We selected the Daubechies 4 (db4) wavelet, a widely used member of the Daubechies family of orthogonal wavelets.

Daubechies wavelets are particularly effective for analyzing non-stationary signals due to their compact support and vanishing moments. Compact support ensures temporal localization and detection of signal features such as sudden spikes or drops. The number of vanishing moments determines the wavelet’s ability to represent smooth trends \cite{WHITE20011419}. Specifically, a Daubechies wavelet with $M$ vanishing moments (dbM) is orthogonal to all polynomials of degree $M - 1$ or less. Thus, db4—being orthogonal to cubic and lower-degree polynomials—effectively captures and preserves cubic trends in the approximation coefficients, while non-polynomial components (including noise and sharp transitions) appear in the detail coefficients.  That means if part of the signal behaves like a cubic polynomial, that portion will primarily appear in the approximation, while deviations from the trend (e.g., noise or abrupt shifts) will be captured in the detail coefficients.

We implemented the decomposition using the \texttt{pywt.wavedec} function from the \texttt{PyWavelets} library in \textsf{Python}, which applies the db4 filters to perform efficient multilevel decomposition. 

\subsection{Sample-Type Comparison Methodology}
We decomposed the normalized wastewater concentration time series for each city and method using the db4 wavelet with $L$ levels of decomposition. This process yielded $L$ approximation coefficient vectors and $L$ detail coefficient vectors, each associated with a distinct frequency band. The length of the coefficients at each level $l$ follows a dyadic scaling pattern: $N / 2^l $, where $N$ is the length of the original time series and  $l \in \{1, \ldots, L\}$.

\subsubsection{Progressive Filtering}\label{sec:filtering}
To explore whether differences between sample types were concentrated in specific frequency ranges, we reconstructed progressively smoothed versions of each signal by systematically zeroing the detail coefficients from the highest to the lowest frequency bands. Specifically, for each level $l \in \{1,...,L\}$, we constructed filtered signals $S_l$ by retaining only the detail coefficients from levels $l+1$ to $L$, while setting higher-frequency details coefficients (levels $1$ to $l$) to zero:

$$S_l= \text{IDWT}(cA_L, \underbrace{0, \ldots, 0}_{l\text{ terms}}, cD_{l+1},\ldots cD_{L})$$
where $\text{IDWT}(\cdot)$ denotes the inverse discrete wavelet transform used to reconstruct the time-domain signal from the modified set of coefficients. This progressive filtering approach removes  high-frequency components up to level $l$, increasingly emphasizing longer-term signal trends. 

The set of filtered signals $\{S_l\}_{l=1}^L$ allows us to assess which frequency components contribute most to differences between sample types.

%In practice, many studies do use Ward linkage with correlation‐based distances when clustering by shape.
\subsubsection{Clustering Analysis}
To identify the level at which methodological noise is minimized, we performed agglomerative hierarchical clustering on the filtered signals $\{S_l\}_{l=1}^L$, using correlation distance as a similarity metric and agglomerative clustering with Ward's linkage.
Ward’s linkage with correlation distance is widely used in genomics for shape-based clustering of standardized expression profiles \cite{Leupin2006,hovestadt2019resolving,talbot2005gene}. In those studies, expression profiles are typically normalized using Z-scores, and clustering is performed to identify genes with similar temporal or conditional expression trajectories, an approach conceptually aligned with our objective of identifying wastewater signals that reflect common epidemiological trends.

%analogous to our goal of grouping signals by shared epidemiological trends.
By examining how clustering patterns changed across ${S_l}$, we identified the smallest smoothing level $l^*$ at which influent and solids samples from the same city began to cluster together. We interpret this transition as the scale at which methodological variation is filtered out, revealing the shared epidemiological trends across sample types.

\section{Results}
\subsection{Wavelet Decomposition}

We applied a three-level discrete wavelet decomposition using the db4 wavelet to the normalized wastewater concentration from influent and settled solids samples collected in the cities of Esparto, Los Banos, Turlock, Woodland, and Winters. This decomposition produced three sets of detail coefficients ($cD_1$, $cD_2$, $cD_3$), corresponding to increasingly lower-frequency components, and one set of approximation coefficients ($cA_3$), which captures the coarsest-scale trends in the data.

\begin{figure}[H]
    \centering
    \includegraphics[width=0.9\linewidth]{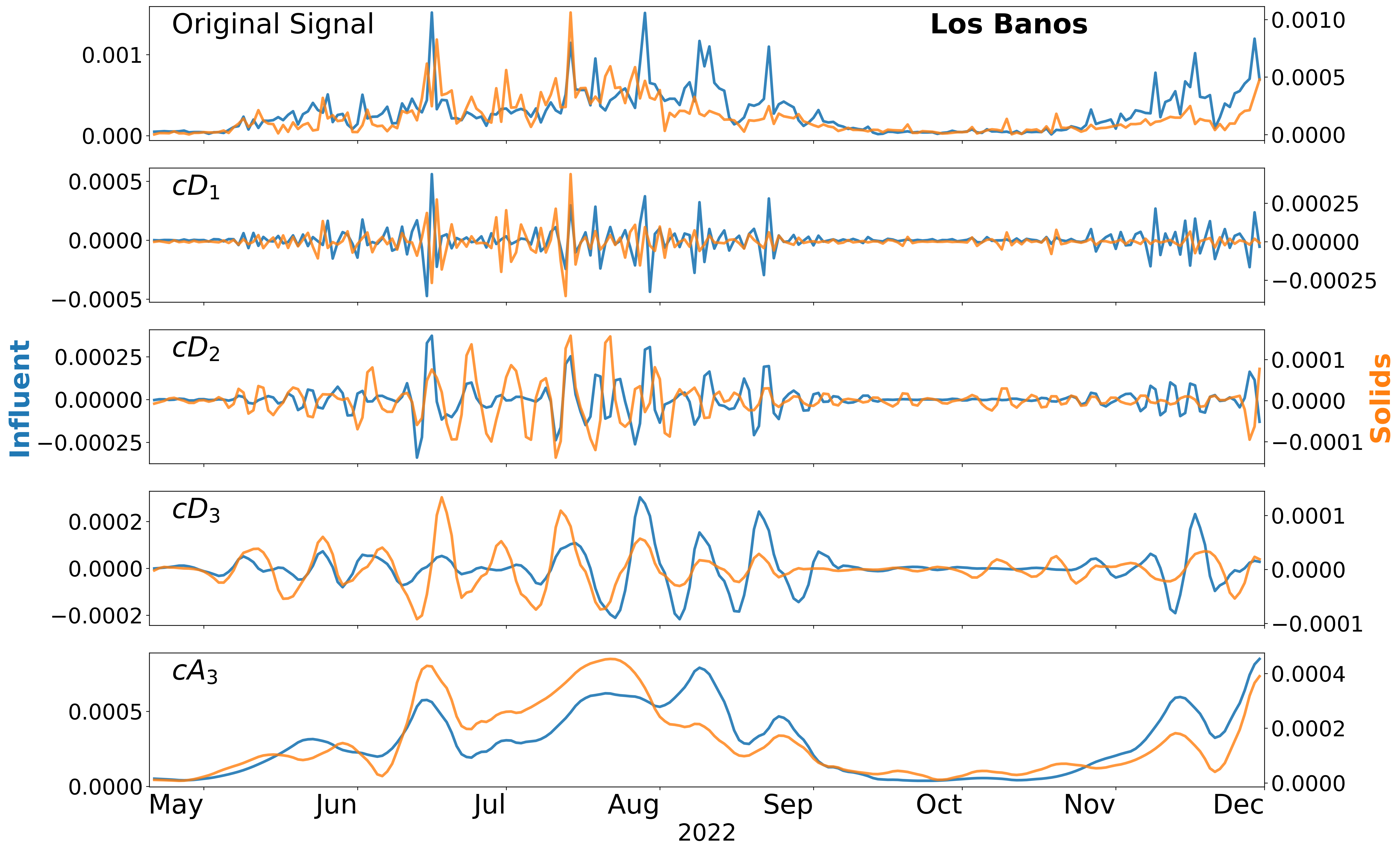}
\caption{Discrete wavelet decomposition of SARS-CoV-2 signals from wastewater influent and settled solids samples in Los Banos. The top panel shows the original time series. The subsequent panels display reconstructed components at each decomposition level: detail coefficients ($cD_1$, $cD_2$, $cD_3$) and the approximation coefficient ($cA_3$).} \label{fig:wavelet_coeff_comp_Los_Banos}
\end{figure}

\begin{figure}[H]
    \centering
   \includegraphics[width=0.54\linewidth]   {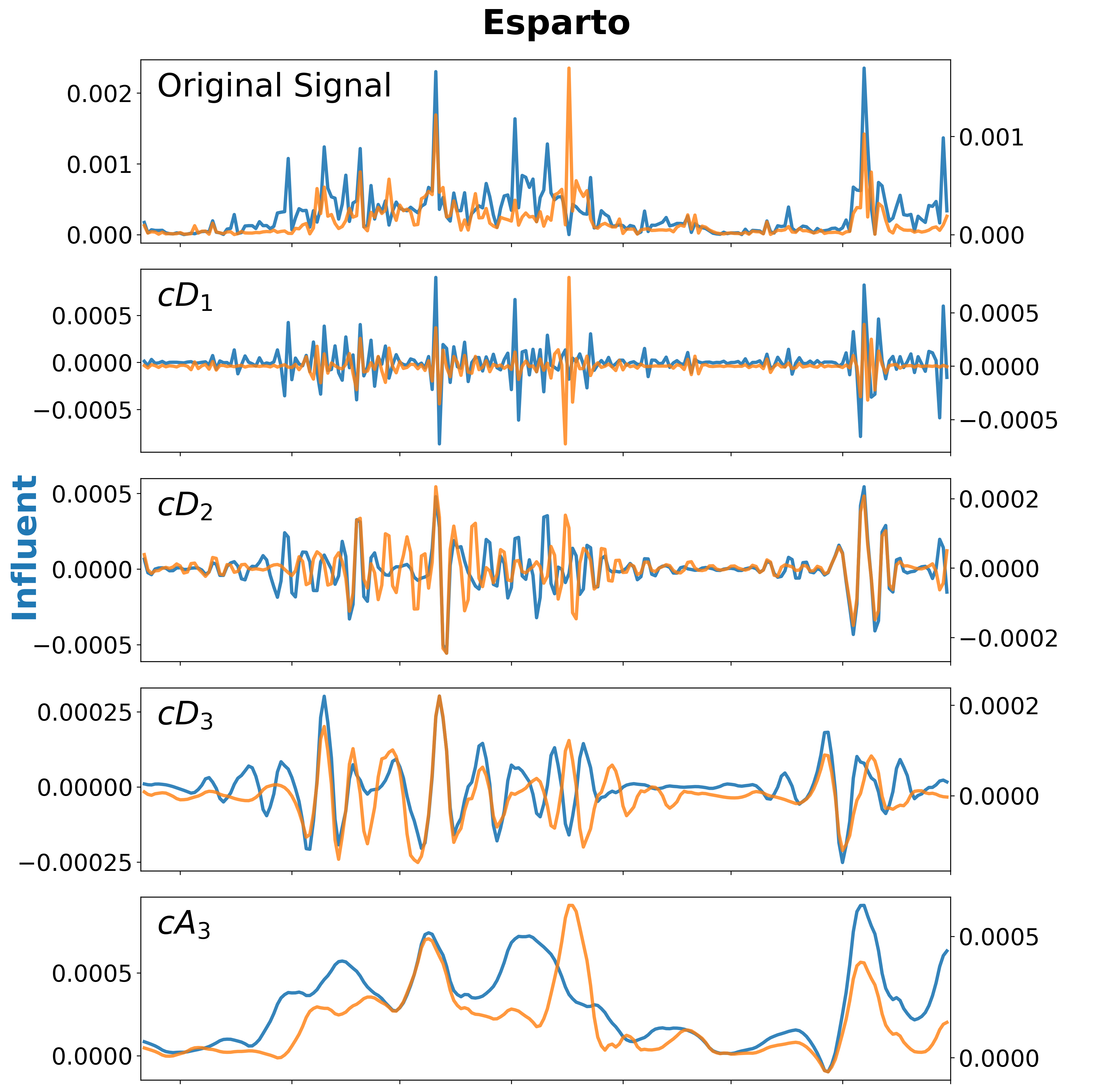}\includegraphics[width=0.54\linewidth]{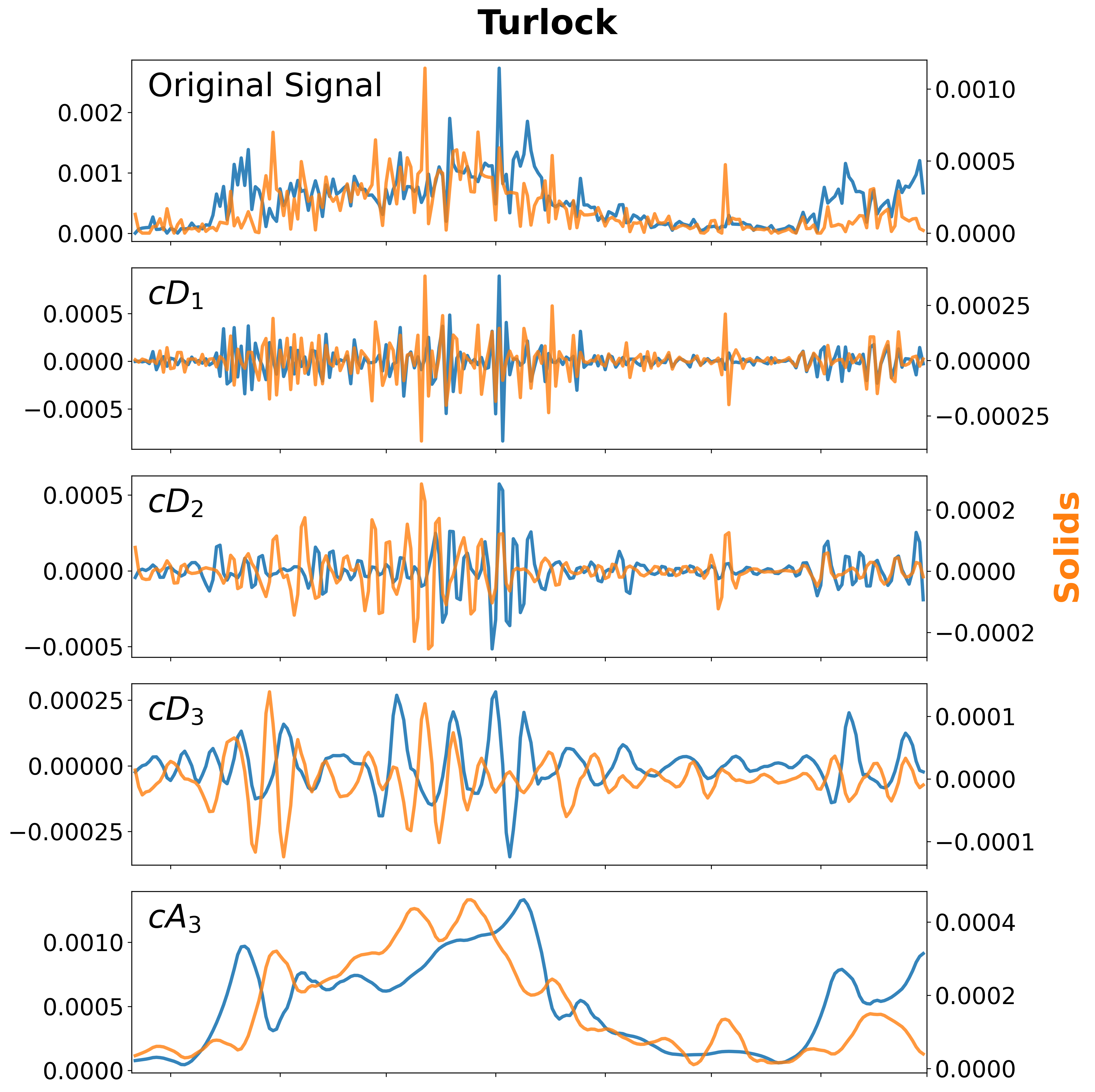}
   \includegraphics[width=0.54\linewidth]{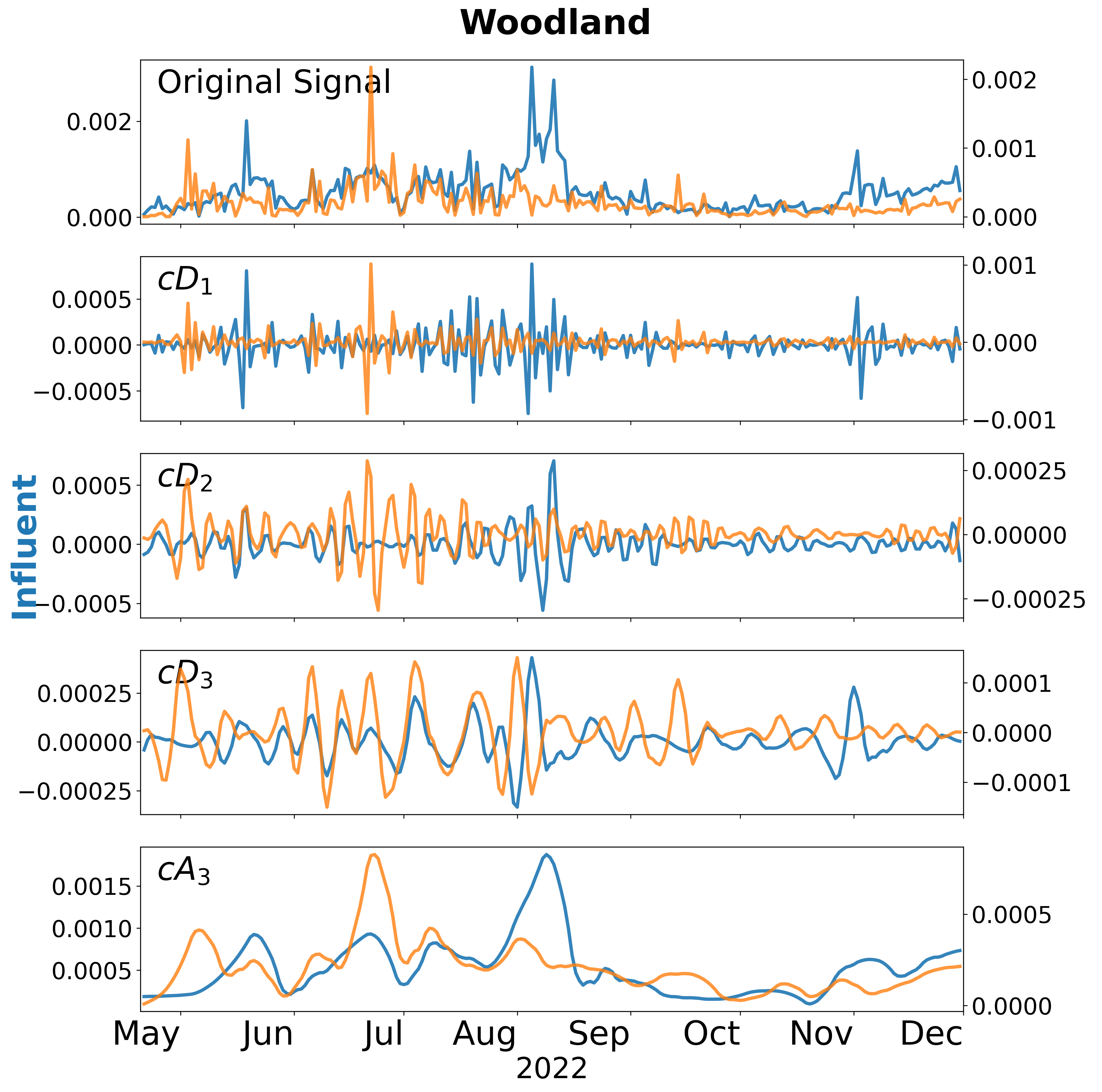}\includegraphics[width=0.54\linewidth]{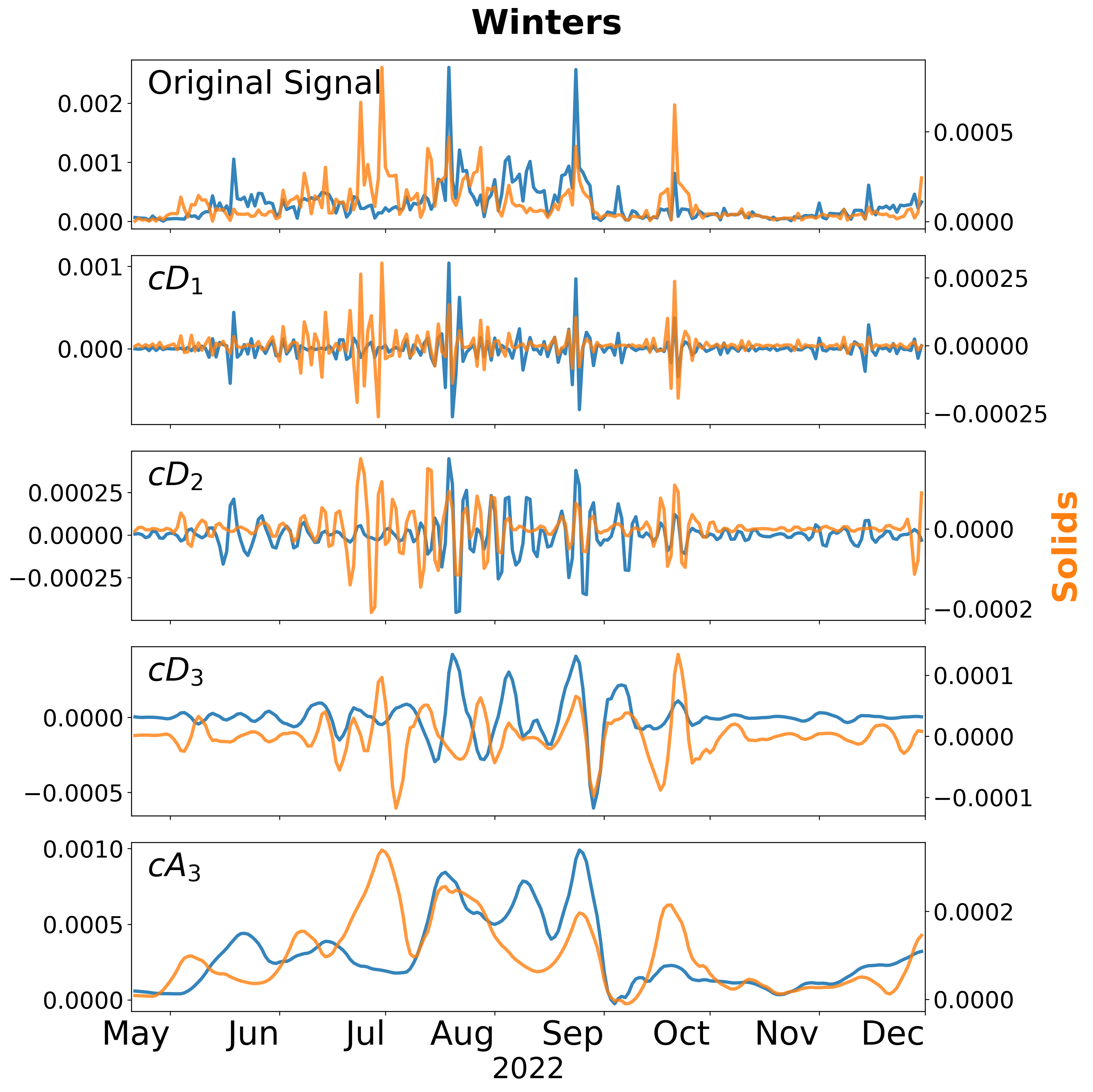}
    \caption{Discrete wavelet decomposition of SARS-CoV-2 signals from wastewater influent and settled solids samples in Esparto, Turlock, Woodland, and Winters. For each city, the top panel shows the original time series for both sample types, followed by panels displaying the reconstructed components for each wavelet level: detail coefficients ($cD_1$, $cD_2$, $cD_3$) and the approximation coefficient ($cA_3$). }
    \label{fig:wavelet_coeff_comp_all}
\end{figure}

To analyze temporal patterns across frequency bands, we reconstructed individual components using the inverse discrete wavelet transform. These reconstructions allow us to compare the detail and approximation components at the original temporal scale across sample types and decomposition levels. The reconstructed coefficients $cD_1$, $cD_2$, $cD_3$, $cA_3$ are plotted alongside the original time series in Figures \ref{fig:wavelet_coeff_comp_Los_Banos} and
\ref{fig:wavelet_coeff_comp_all}.  The corresponding coefficient estimates are presented in Figures \ref{fig:S_wavelet_coeff_LosBanos}-\ref{fig:S_wavelet_coeff_all} of the Supplementary Material (SM).

%Qualitatively, these component reconstructions reveal the type of information captured at each level of decomposition. The approximation component ($cA_3$) captures the smooth, underlying trend of the signal, while the detail components ($cD_1-cD_3$) reflect finer-scale fluctuations, including noise and transient events. 
In Figures \ref{fig:wavelet_coeff_comp_Los_Banos}-\ref{fig:wavelet_coeff_comp_all}, we observed larger discrepancies between influent and solids signals within the high-frequency detail components ($cD_1, cD_2,cD_3$), indicating that methodological differences between sample types are more pronounced at shorter temporal scales. In contrast, the approximation component showed a strong concordance across sample types within each city, suggesting that the core epidemiological signal is reliably preserved between influent and settled solids samples.

\begin{figure}[H]
    \centering
    \includegraphics[width=0.9\linewidth]{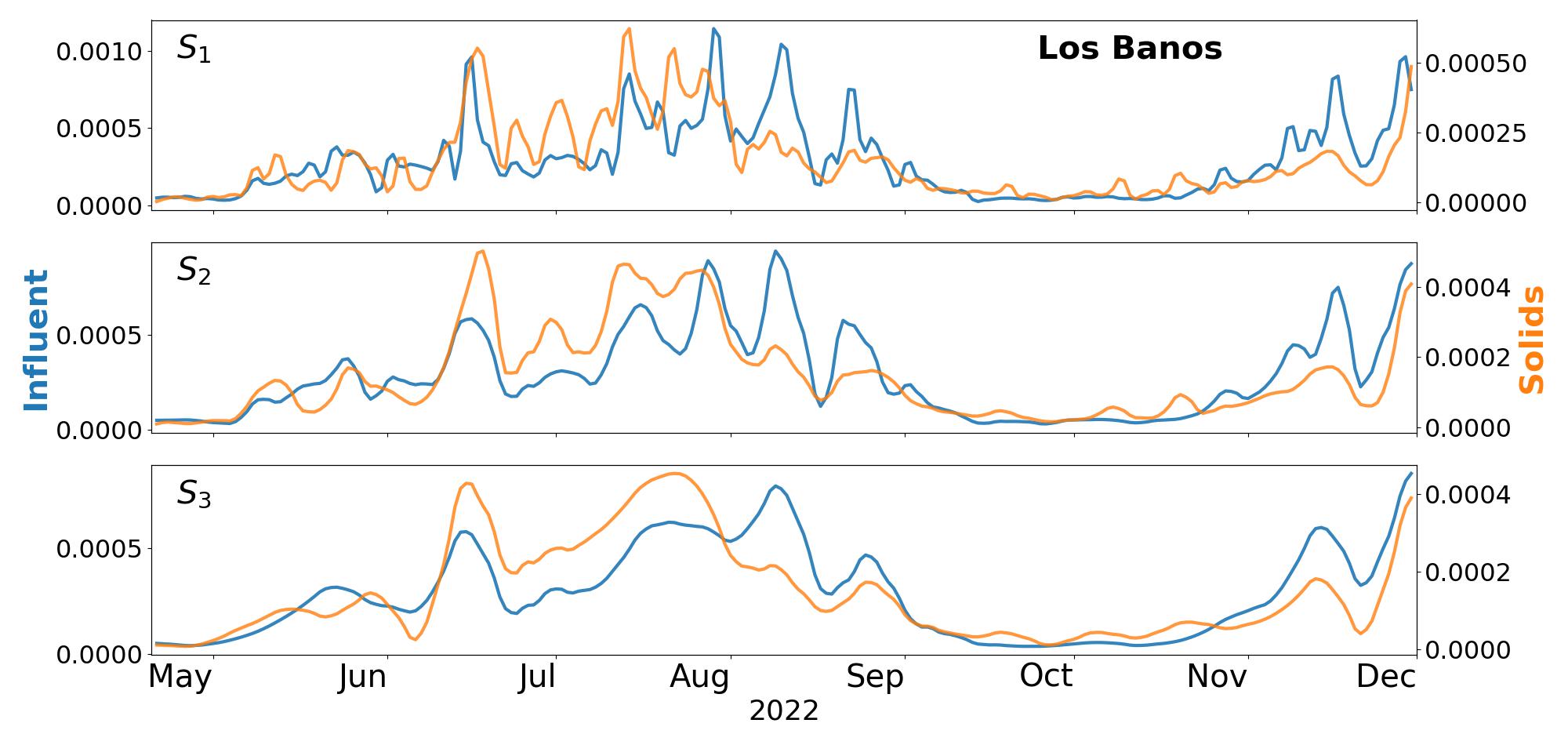}
    \caption{Progressively smoothed SARS-CoV-2 signals from wastewater influent and settled solids wastewater samples in Los Banos. Each panel shows a filtered $S_l$, for $l = 1, 2, 3$, obtained by removing the first $l$ levels of detail coefficients using inverse discrete wavelet transform. Influent and solids signals are overlaid in each panel to highlight their alignment at different levels of smoothing: $S_1$ (top), $S_2$ (middle), and $S_3$(bottom).}

    \label{fig:rec_Los_Banos}
\end{figure}

\begin{figure}[h!]
    \centering
\includegraphics[width=0.53\linewidth]{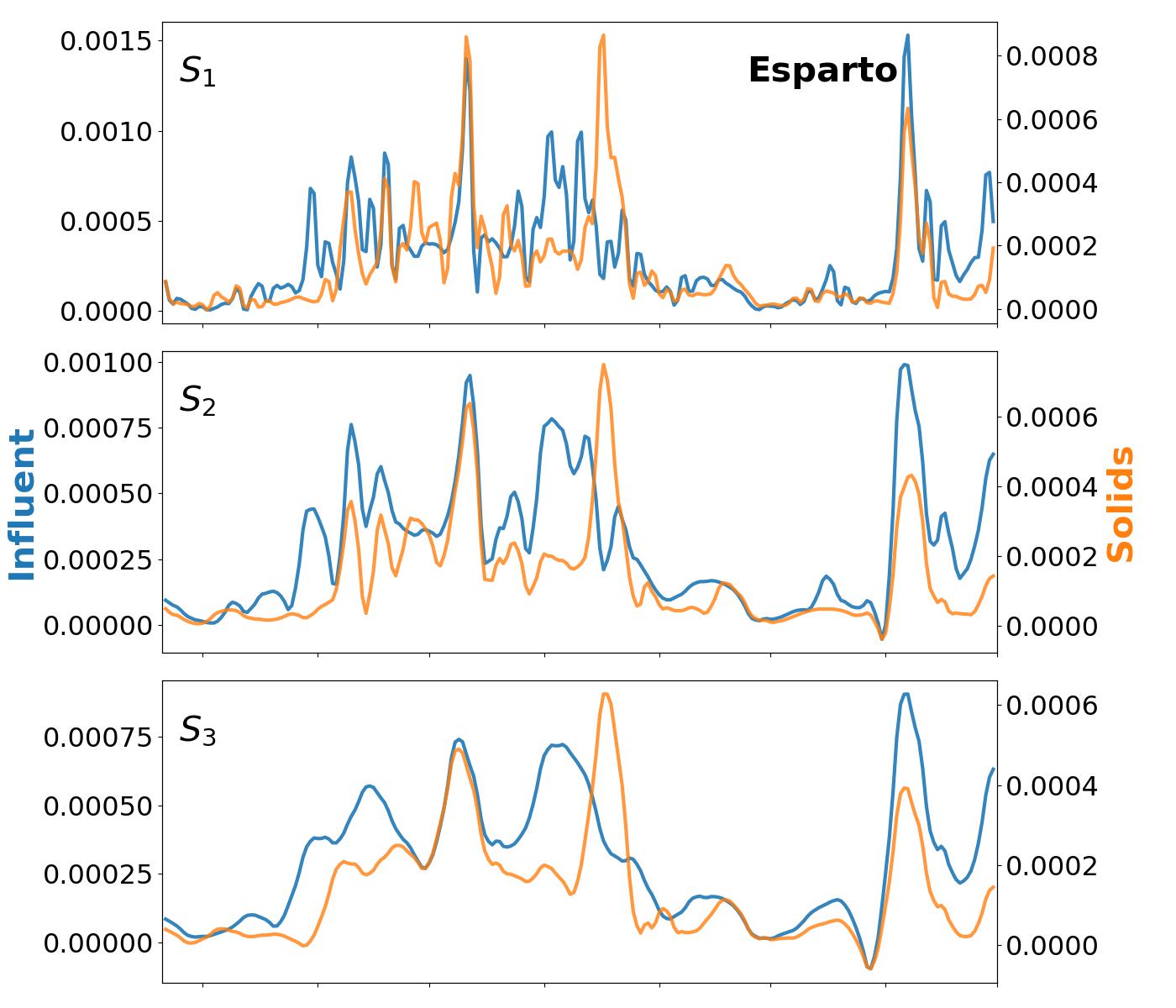}\includegraphics[width=0.53\linewidth]{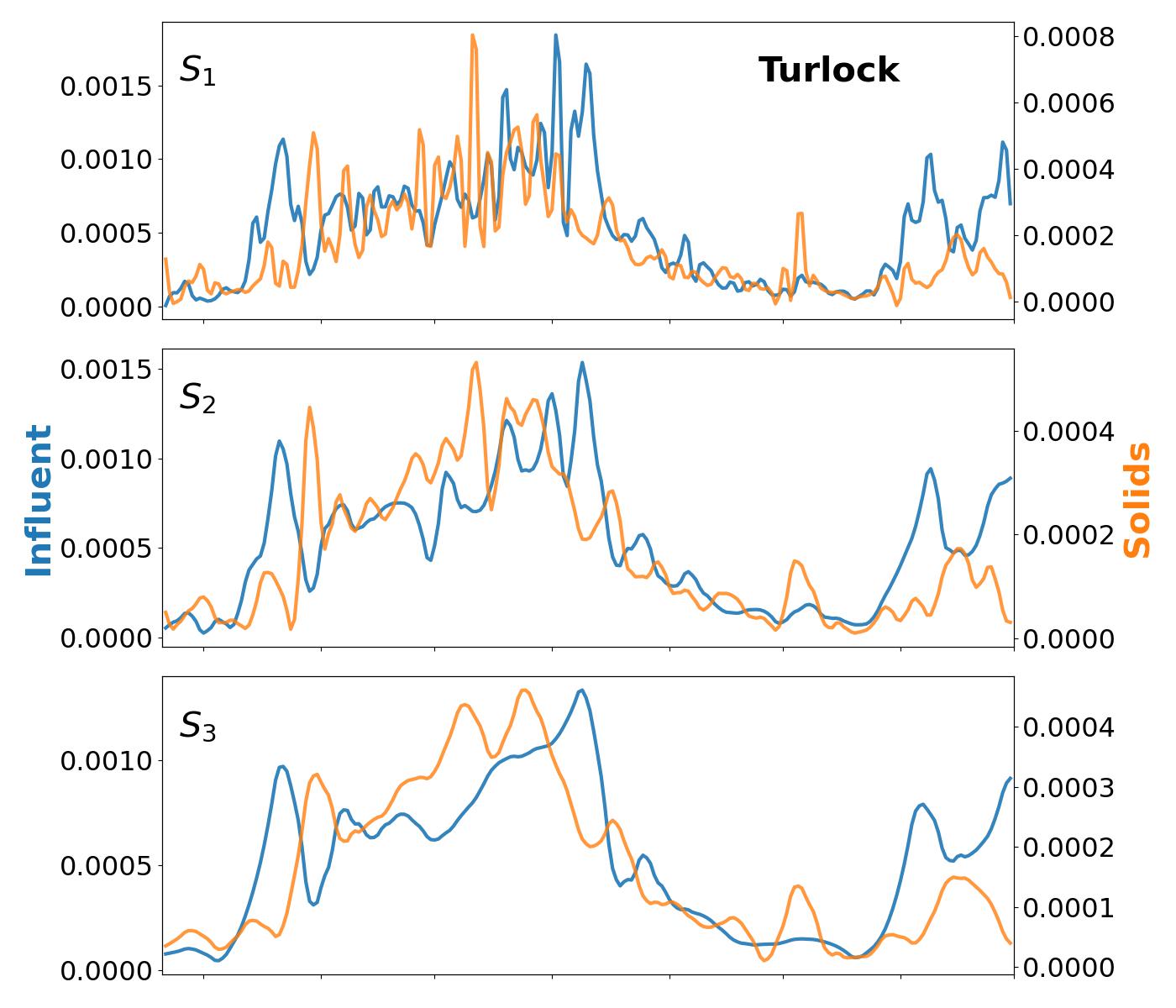}
\includegraphics[width=0.53\linewidth]{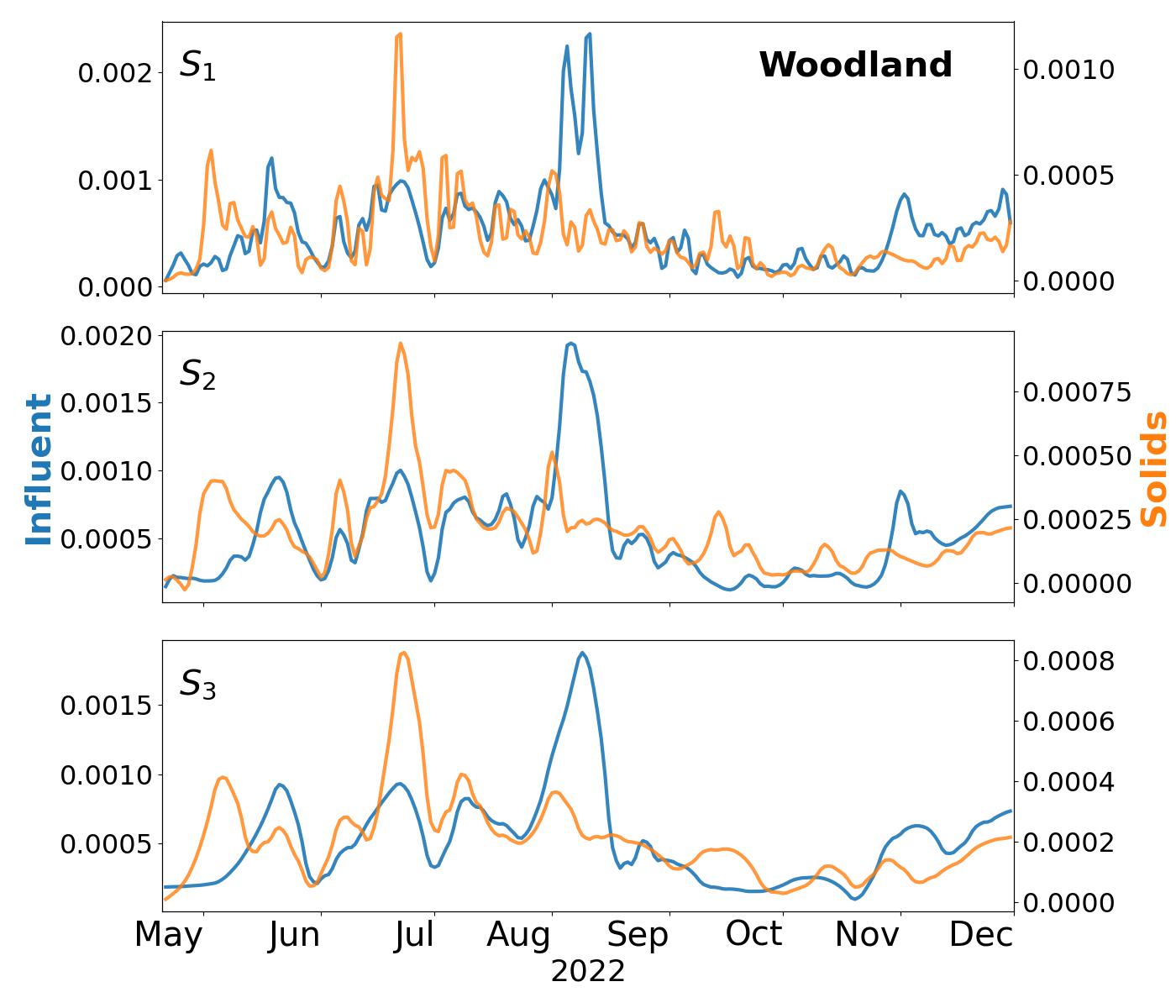}\includegraphics[width=0.53\linewidth]{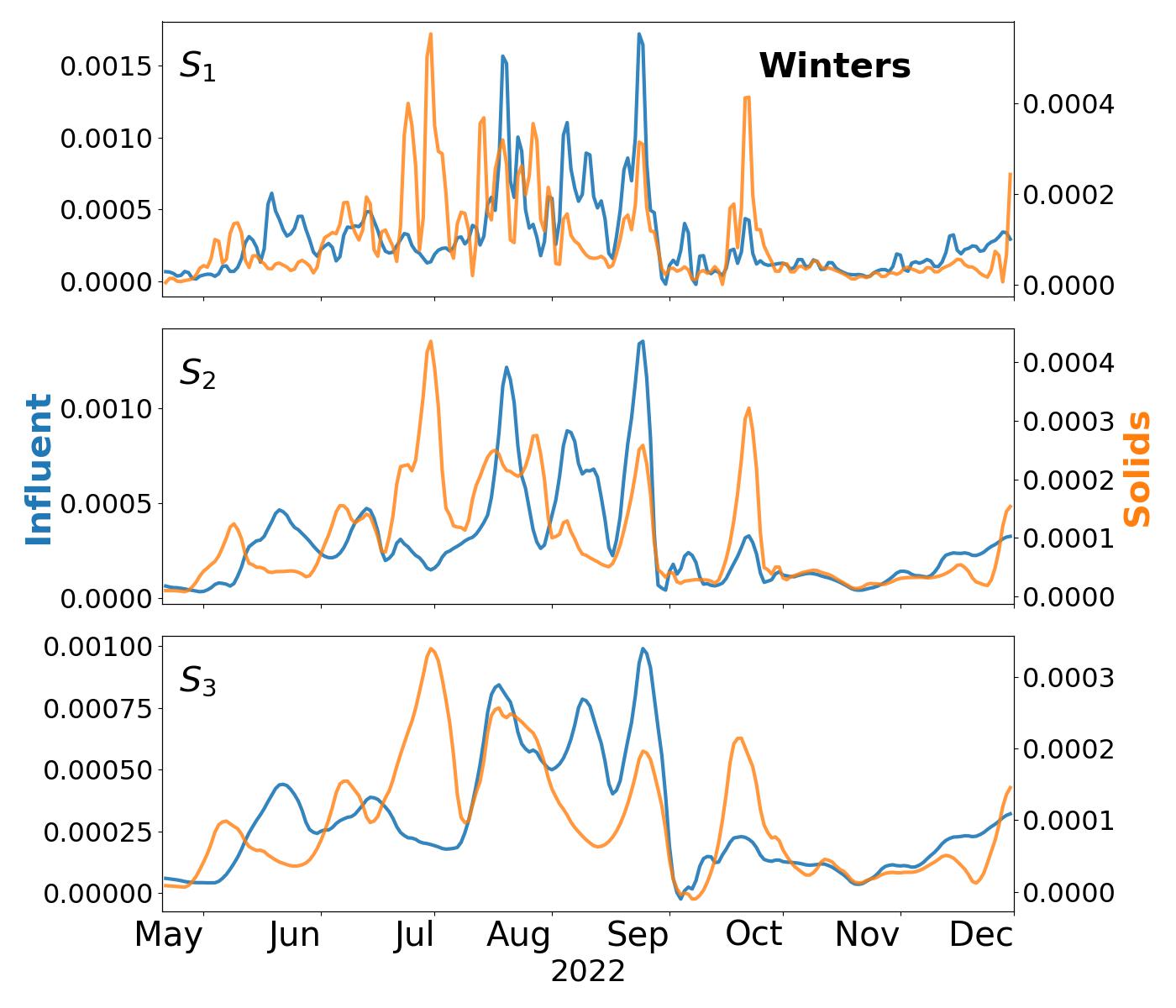}

\caption{Progressively smoothed SARS-CoV-2 signals from wastewater influent and settled solids samples in Esparto, Turlock, Woodland, and Winters. Each subplot corresponds to one city and displays three filtered signals $S_l$, for $l = 1, 2, 3$, obtained by removing the first $l$ levels of detail coefficients using inverse discrete wavelet transform. Within each subplot, influent and solids signals are overlaid for each level of smoothing to visualize how alignment improves as high-frequency fluctuations are removed: $S_1$, $S_2$, and $S_3$.}
\label{fig:rec_all}
\end{figure}

To assess whether high-frequency fluctuations obscure shared trends across sample types, we progressively filtered the signals by removing increasing levels of detail coefficients, as described in Section \ref{sec:filtering}. This process produced a series of smoothed signals, denoted $S_l$ (for $l = 1, 2, 3$), each retaining progressively more low-frequency content (Figures \ref{fig:rec_Los_Banos} and \ref{fig:rec_all}). 
These refined smoothed signals were subsequently used in clustering analyses to evaluate whether filtering out high-frequency noise enhanced the concordance between influent and settled solids signals within each city.

%All panels share the same time and amplitude scales, with influent and sludge signals overlaid to facilitate direct multi-scale comparison.

%All component panels maintain consistent time and amplitude scales, with influent and sludge signals overlaid to highlight differences and similarities across frequency bands.

\subsection{Cluster Analysis of Filtered Signals}
To assess the impact of wavelet-based filtering, we performed hierarchical clustering on both raw and filtered time series for each city and sample type (influent and solids). Each time series was normalized by its maximum value and standardized to emphasize relative signal patterns over absolute magnitudes. Clustering was conducted using the \texttt{scipy.cluster.hierarchy} module in \textsf{Python}, with Ward’s linkage method and correlation distance as the dissimilarity metric. Ward’s method merges clusters iteratively by minimizing within-cluster variance, producing compact and well-separated groups. 

\begin{figure}[h!]
    \centering
\includegraphics[width=1\linewidth]{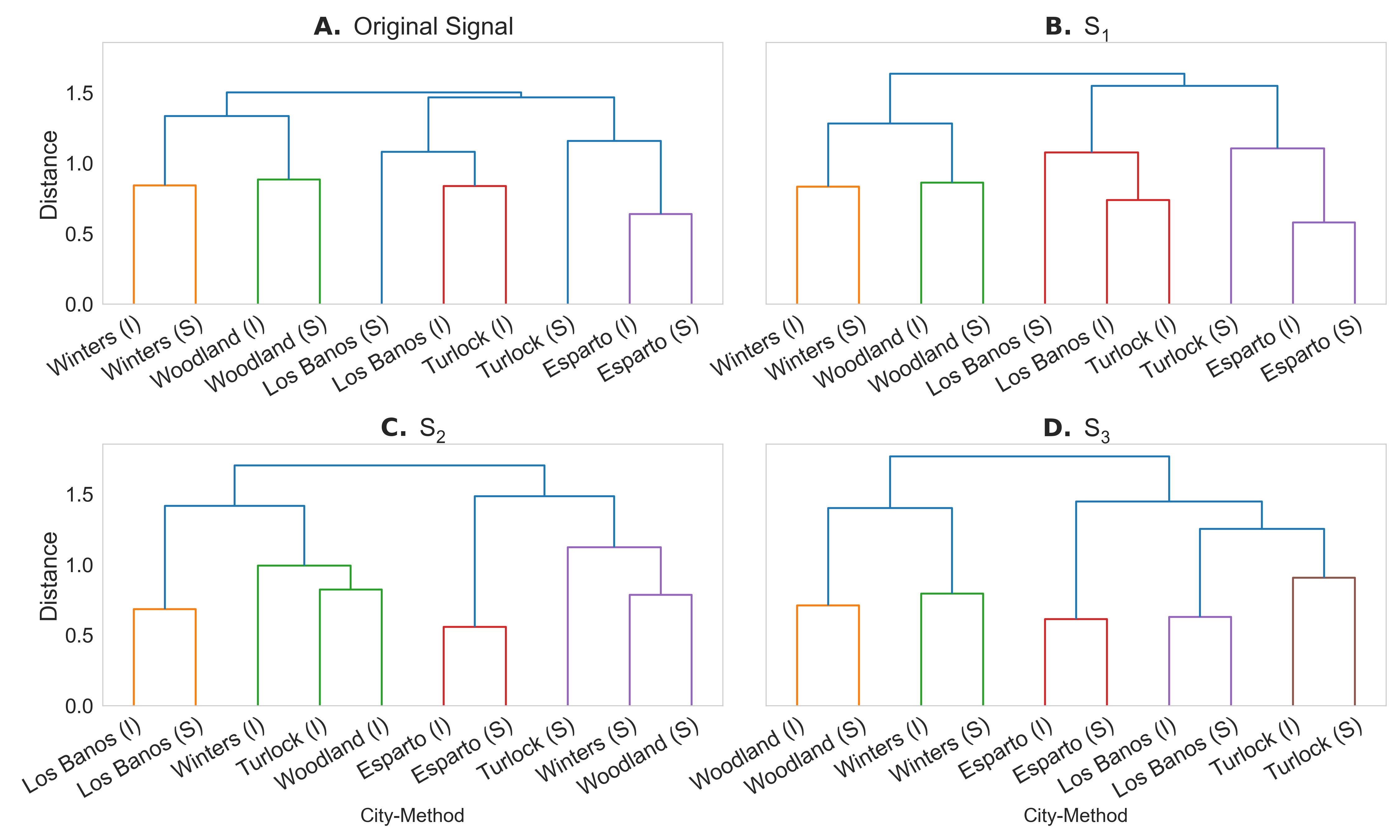}
    \caption{Hierarchical clustering dendrograms of SARS-CoV-2 signals across cities and sample types. Each panel shows clustering results for Los Banos, Turlock, Winter, Woodland, and Esparto, based on different versions of the time series: {\bf (A)} original (unfiltered) signals, reconstructed signals  {\bf (B)} S1 (retaining $cD_2$ and $cD_3$), {\bf (C)} $S_2$ (retaining $cD_3$), and {\bf (D)} $S_3$ (retaining only the $cA_3$ approximation). On the x-axis of each dendrogram, sample labels are denoted as City(I) for influent samples and City(S) for solids samples. Clustering was performed using Ward’s linkage and correlation distance. City-specific grouping becomes more distinct as higher-frequency components are removed, with $S_3$ showing clear separation by city across sample types.}
    \label{fig:clusters}
\end{figure}

As shown in Figure \ref{fig:clusters}, clustering analysis of the unfiltered (raw) signals did not reveal distinct groupings by city, indicating a lack of clear city-specific patterns in the raw data. Similarly, clustering of the reconstructed signals $S_1$ (retaining $cD_2$ and $cD_3$)  and $S_2$ (retaining $cD_3$) still resulted in mixing of cities and sample types. This suggests that methodological and sample-specific noise continued to dominate the signal even after partial filtering.  In contrast, clustering $S_3$ signals (retaining only $cA_3$) produced distinct city-level clusters, where influent and solids samples from the same city grouped together. This result supports the hypothesis that high-frequency components ($cD_1$–$cD_3$) capture predominantly noise, while low-frequency component ($cA_3$) preserves shared epidemiological trends across sample types.

To further investigate these patterns, we also clustered the wavelet coefficients directly. Clustering the detail coefficients ($cD_1-cD_3$) yielded no separation by city. In contrast, clustering $cA_3$ coefficients alone replicated the city-level separation observed in $S_3$. These additional dendrograms are included in the Supplemental Material (SM) Section (Figure \ref{fig:S_clusters}). This finding supports the hypothesis that, although wastewater signal variability is influenced by methodological and environmental noise, these effects are predominantly confined to specific frequency bands. 

\begin{figure}[H]
    \centering
    \includegraphics[width=0.52\linewidth]{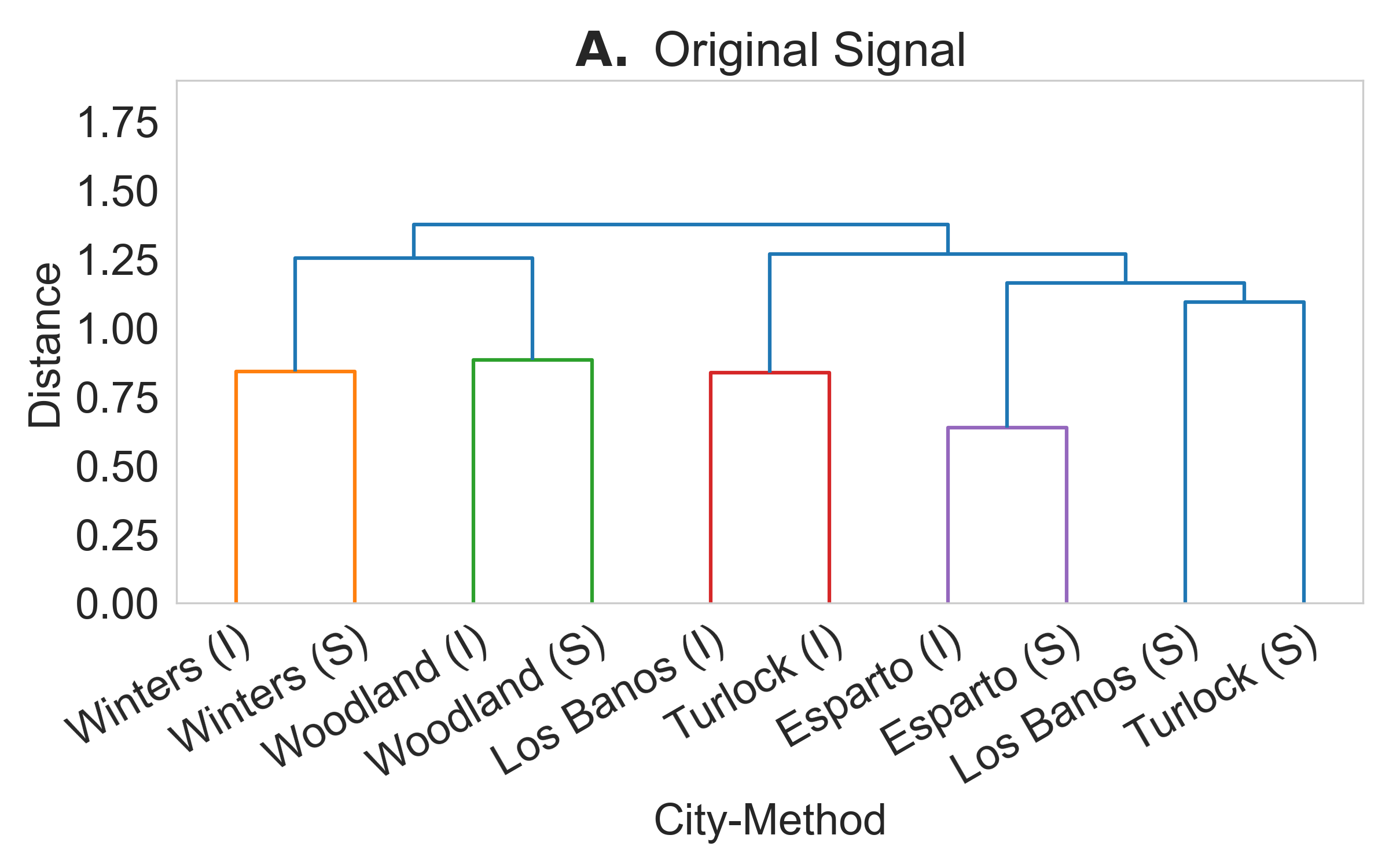}\includegraphics[width=0.52\linewidth]{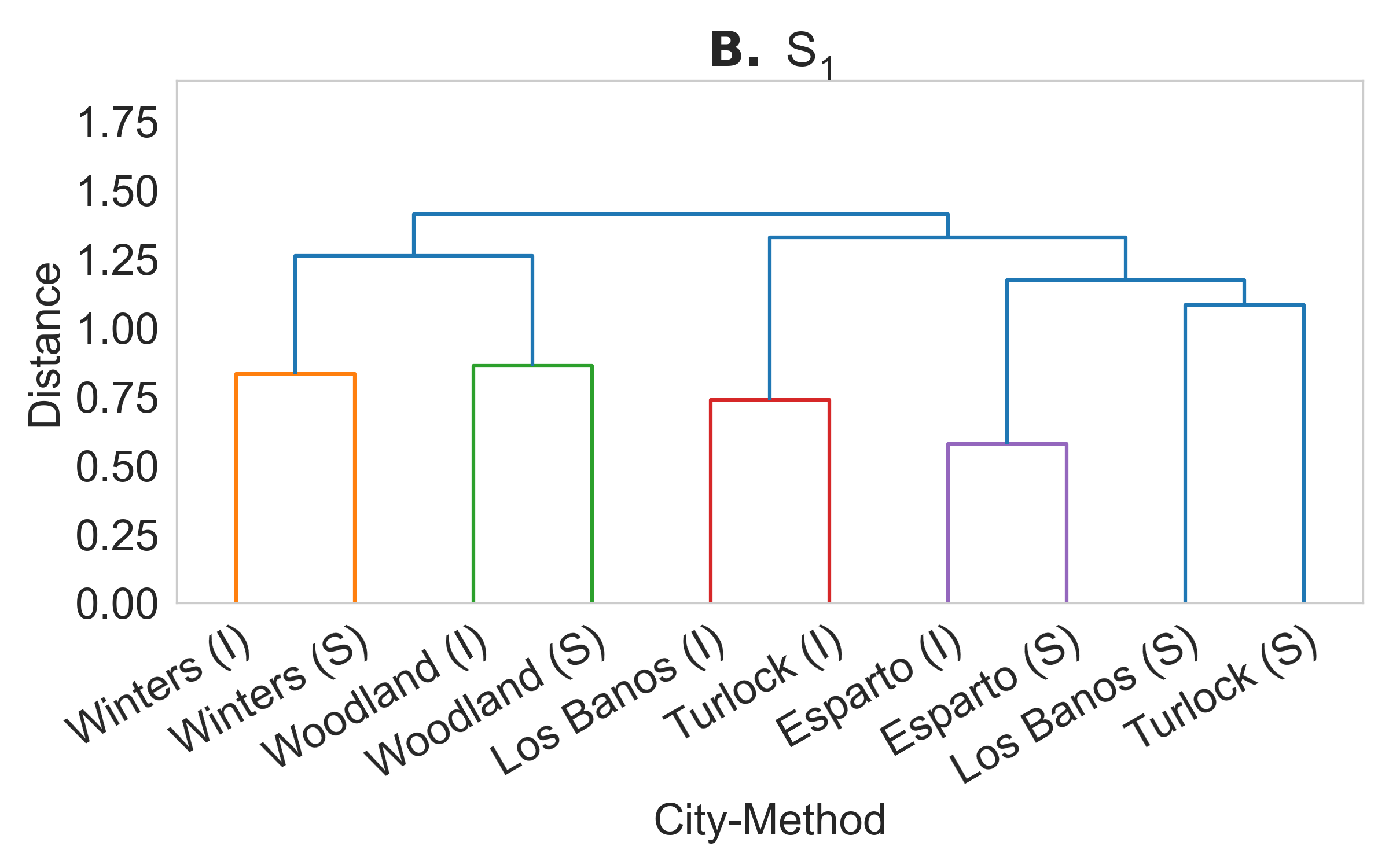}
    \includegraphics[width=0.52\linewidth]{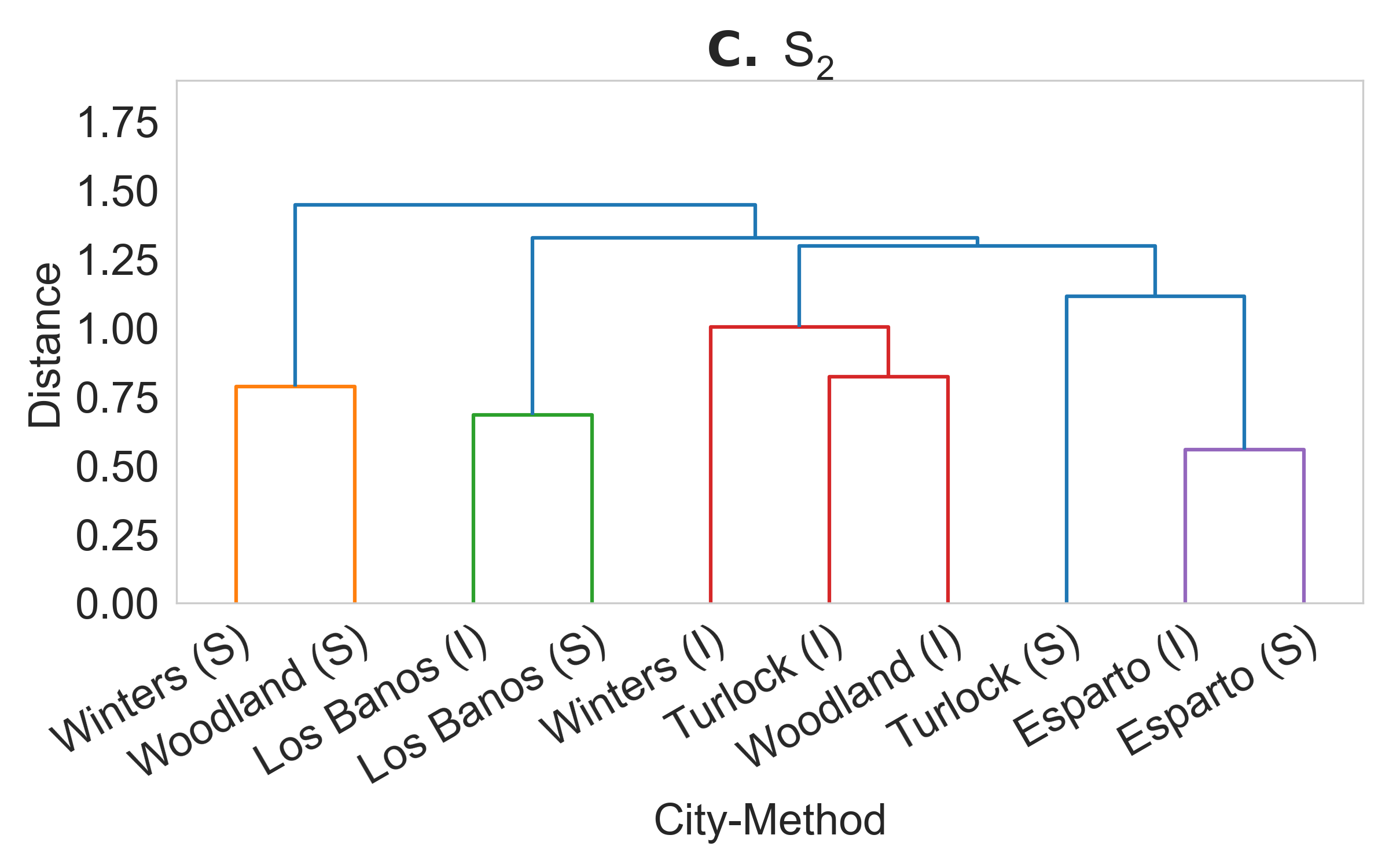}\includegraphics[width=0.52\linewidth]{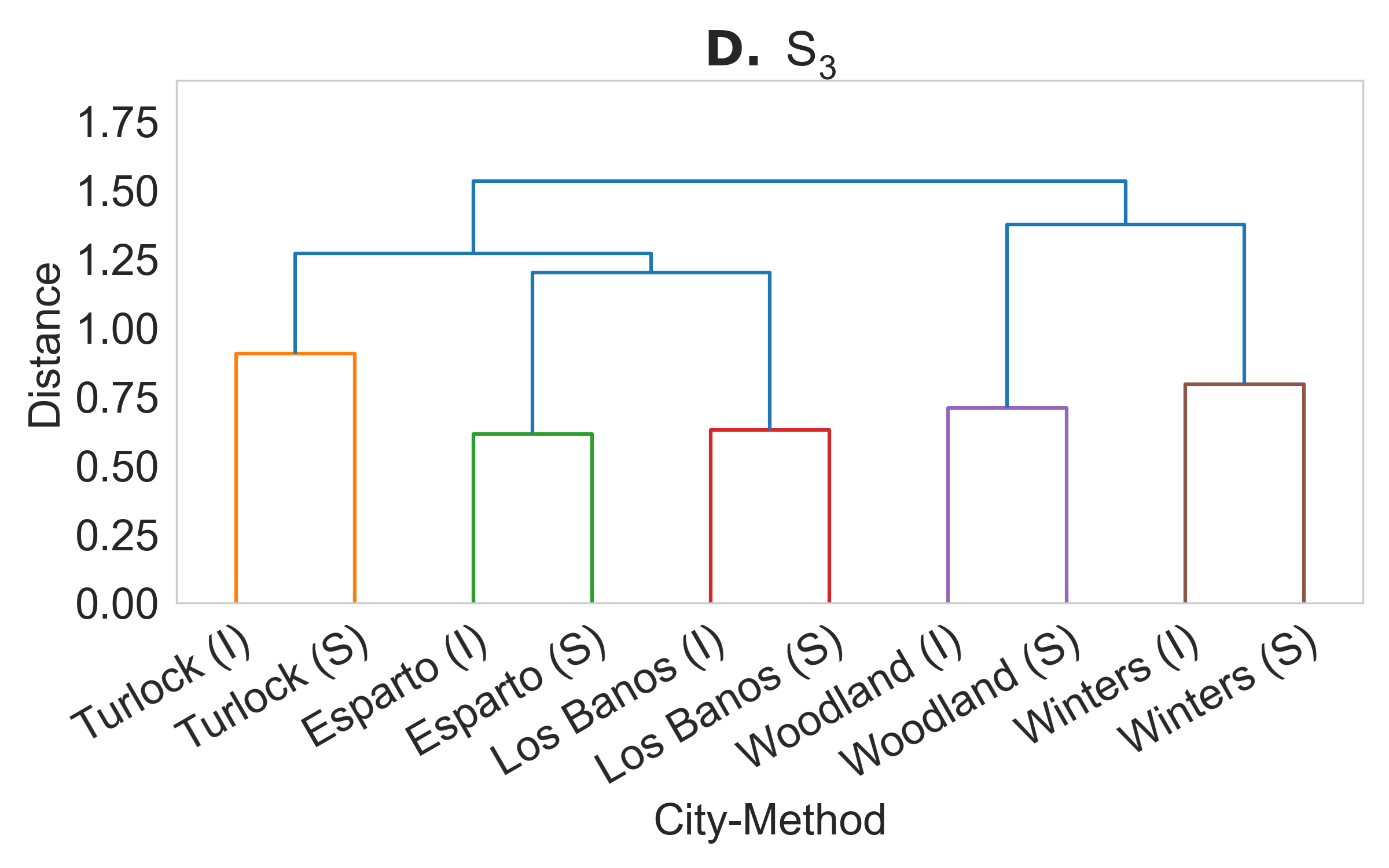}
    \includegraphics[width=0.52\linewidth]{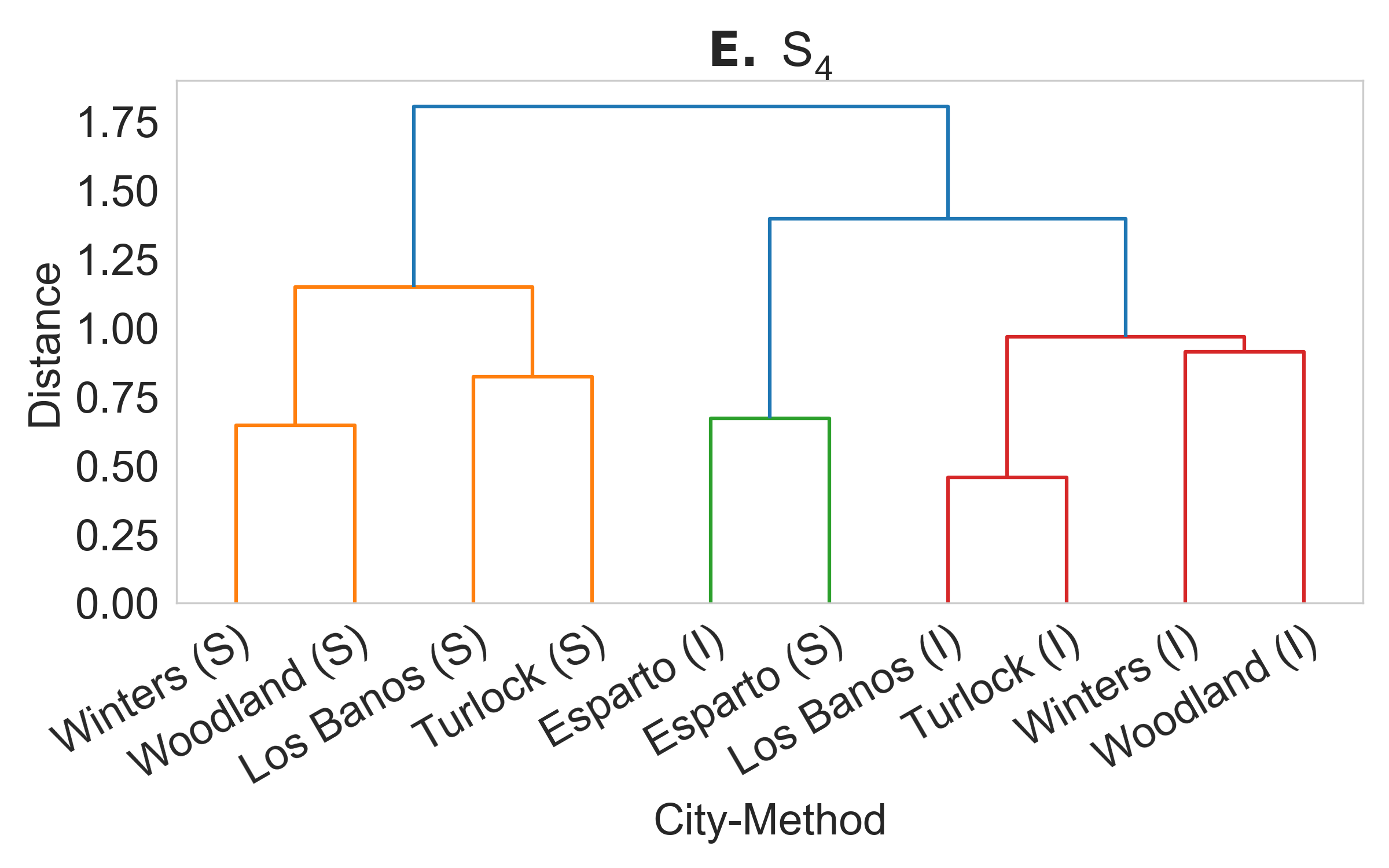}

\caption{Hierarchical clustering dendrograms of SARS-CoV-2 signals from wastewater samples, reconstructed after a four-level wavelet decomposition.  Each panel shows clustering results for Los Banos, Turlock, Winter, Woodland, and Esparto, based on different versions of the time series: {\bf (A)} original (unfiltered) signals, reconstructed signals  {\bf (B)} S1 (retaining $cD_2,cD_3,cD_4$), {\bf (C)} $S_2$ (retaining $cD_3,cD_4$), , {\bf (D)}  $S_3$ (retaining $cD_3,cD_4$), {\bf (D)} $S_4$ (retaining only the $cA_4$ approximation). On the x-axis of each dendrogram, sample labels are denoted as City(I) for influent samples and City(S) for solids samples. City-specific grouping becomes more distinct as higher-frequency components are removed, with $S_3$ showing clear separation by city across sample types.  }
\label{fig:S_cluster_level4}
\end{figure}

We extended this analysis to a four-level decomposition ($L=4$), producing four sets of detail coefficients ($cD_1$–$cD_4$) and a level-4 approximation coefficient ($cA_4$). Signals reconstructed from these components, denoted $S_1$ through $S_4$, were again subjected to clustering. We observed that clustering $S_4$—which retains only the coarsest approximation component $cA_4$—failed to group samples by city, resulting in mixing across cities and sample types (see Figure \ref{fig:S_cluster_level4}, SM). This suggests that at level 4, critical epidemiological signals may be lost due to excessive smoothing. Therefore, level 3 decomposition appears to provide an optimal trade-off between noise reduction and signal preservation.

It is also important to note that the use of a db4 wavelet implies an assumption of local signal smoothness up to third-order polynomials. This makes the detail coefficients particularly sensitive to abrupt, high-order fluctuations—precisely the type of localized, sharp variability introduced by laboratory or environmental noise. The approximation coefficients, on the other hand, approximate piecewise cubic trends, analogous to cubic B-splines but without requiring fixed knot spacing or global stationarity assumptions.

Unlike conventional smoothing techniques—such as LOESS (local regression with fixed bandwidth), nonparametric kernel regression (dependent on predefined kernel width), Kalman filters (requiring state-space assumptions), cubic B-splines (fixed knot spacing), or moving averages (fixed window size)—wavelet decomposition adaptively localizes features in both time and frequency. Traditional methods impose a single smoothing scale through user-defined parameters, leading to critical limitations: LOESS and kernel regression oversmooth abrupt epidemiological shifts, while B-splines struggle with mixed-frequency noise due to fixed knot intervals. Wavelets circumvent these issues by performing multiscale noise-signal separation without prior assumptions about noise structure or trend smoothness. This enables selective attenuation of high-frequency noise while preserving critical local features, such as trend shifts and peaks.

%These low-frequency dynamics captured in the approximation coefficients align closely with epidemiological trends.

\section{Discussion}
Variability in wastewater signals beyond epidemiological drivers arises from a complex interplay of factors, including fluctuations in wastewater flow rates, inconsistencies in sampling, RNA degradation during transport or storage, artifacts introduced during laboratory processing, among others. Decomposition of signals into approximation and detail coefficients allows the localization of sources of uncertainty within defined frequency bands, enabling more precise separation of epidemiological patterns from methodological artifacts and environmental noise.
High-frequency detail coefficients primarily capture short-term fluctuations driven by methodological noise and transient environmental changes, whereas approximation coefficients represent longer-term, smoothed trends that more reliably reflect underlying epidemiological patterns. By distinguishing these frequency bands, we can reduce methodological noise and extract trends that inform public health decision-making and wastewater-based epidemiology more accurately.

This study introduces a novel approach for characterizing and mitigating noise in wastewater signals through multiscale wavelet decomposition. Wastewater data are inherently noisy, reflecting a complex interplay of biological, environmental, and methodological uncertainties. Traditional time-series analyses often fail to disentangle these overlapping sources of variability. By transforming the signal into the time–frequency domain, wavelet decomposition enables the isolation of noise across distinct temporal scales, allowing for a more refined understanding of signal structure and improved extraction of meaningful epidemiological trends. Rather than merely applying a smoothing technique, this work prioritizes disentangling the signal’s structure to distinguish components that convey meaningful epidemiological information from those dominated by noise. 

Our findings indicate that methodological and sample-specific noise is predominantly concentrated in the high-frequency components of the wastewater signal. When these components are retained, they obscure meaningful city-level patterns. However, applying wavelet-based filtering to remove this high-frequency noise unveils clearer, more coherent groupings of signals by geographic location, enhancing our ability to detect underlying epidemiological trends. This has critical implications for regional signal aggregation and inter-city comparisons, as the inclusion of high-frequency noise can obscure underlying epidemiological trends and lead to misleading conclusions. Importantly, this work also highlights several opportunities for future research. 

One promising direction is to examine whether high-frequency components, often dismissed as noise, are systematically linked to external factors such as temperature variation, precipitation, transport time, or specific features of wastewater treatment facilities. Understanding these relationships could refine our interpretation of wastewater signals and improve the reliability of wastewater-based epidemiology. Moreover, refining noise characterization has significant implications for public health surveillance, enhancing the sensitivity and specificity of wastewater monitoring as an early warning system for infectious disease outbreaks and other community health threats. Technological innovations, including advanced wavelet methods and machine learning algorithms, could automate noise detection and correction, fostering more robust and real-time epidemiological insights. 

%In this study, we compared two sample types—influent and solids—across multiple cities. By applying wavelet decomposition to both data signals, we were able to assess the consistency of epidemiological signals across sample types and locations. Our results show that when high-frequency noise is suppressed, influent and solids signals from the same city cluster together, suggesting shared underlying trends despite methodological differences. We demonstrate that wavelet decomposition offers a powerful and nuanced framework for separating noise from meaningful epidemiological signals in wastewater-based surveillance. By leveraging the time–frequency domain, this approach enhances interpretability, supports more accurate regional aggregation, and provides a foundation for methodological standardization across diverse sampling strategies.

\section*{Acknowledgments}
This work and research would not be possible without our wastewater treatment plant (Merced, Modesto, and Davis) and public health partners (Merced County Department of Public Health, Stanislaus County Department of Public Health, Yolo County Department of Public Health, and California Department of Public Health). We also thank the staff of our commercial lab partners. 

This research is supported by a grant from the University of California Alianza MX (UC Alianza MX) 

\bibliography{References.bib}

\newpage
\beginsupplement{S}

\section*{Supplemental Material}

\begin{figure}[H]
\centering
\includegraphics[width=0.8\linewidth]{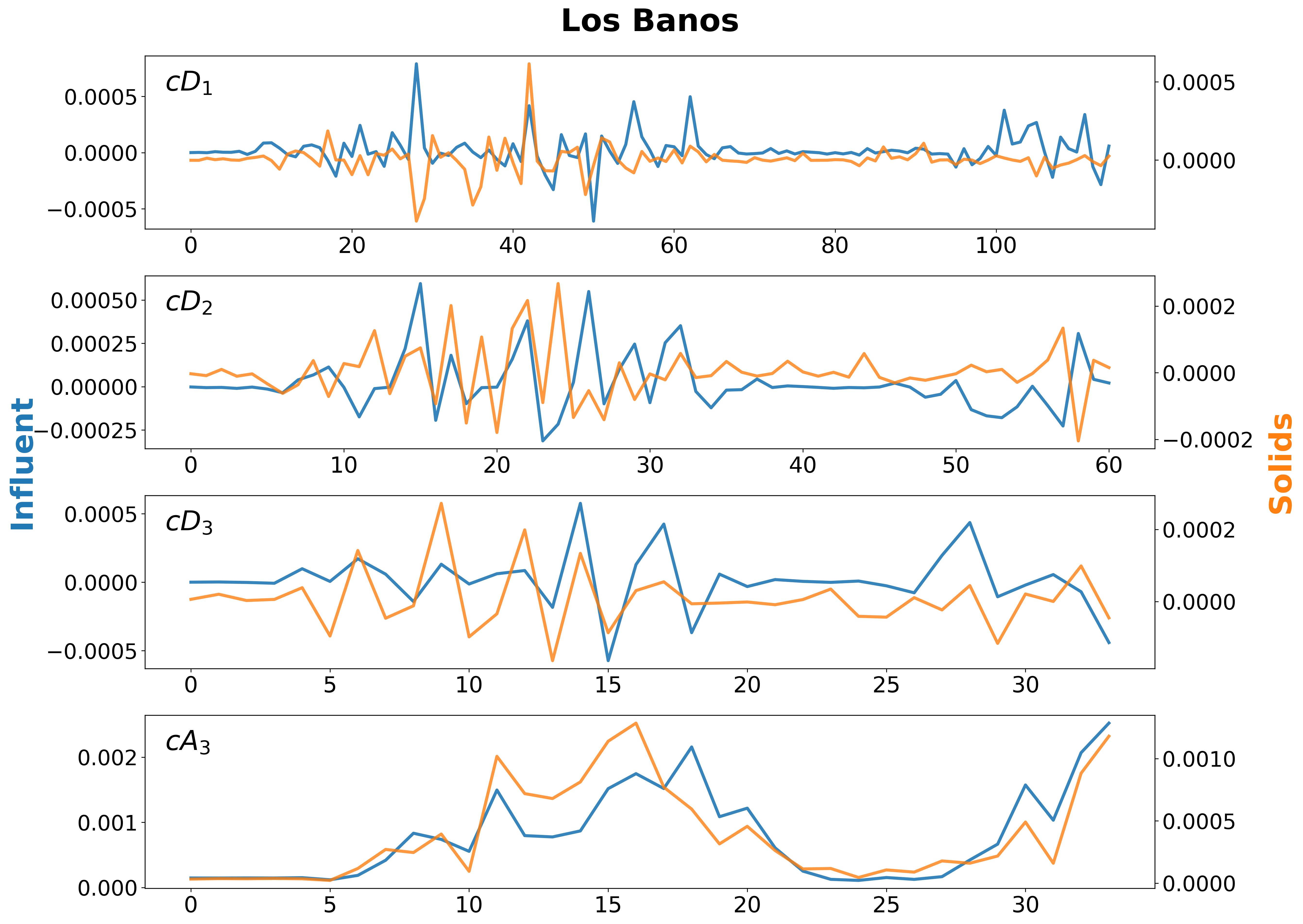}
\caption{Wavelet decomposition coefficients for SARS-CoV-2 concentration signals from wastewater influent and solids samples in Los Banos. The figure displays the detail coefficients ($cD_1$, $cD_2$, $cD_3$) and the approximation coefficient ($cA_3$) obtained from discrete wavelet decomposition. These coefficients represent distinct frequency components of the original time series and are plotted to support the visual comparison of temporal dynamics across sample types and decomposition levels.}
    \label{fig:S_wavelet_coeff_LosBanos}
\end{figure}

\begin{figure}[H]
\centering
\includegraphics[width=0.54\linewidth]{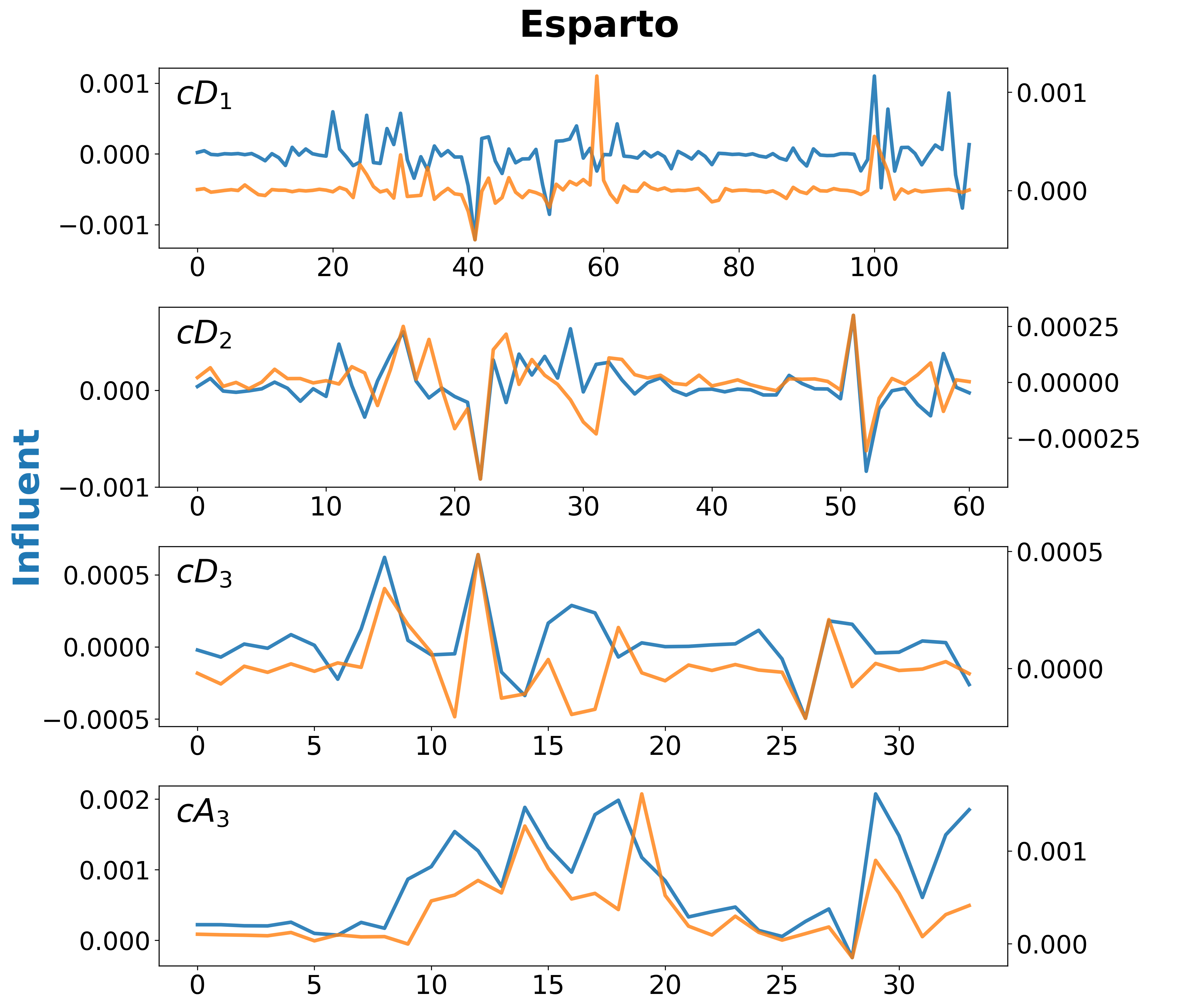}\includegraphics[width=0.54\linewidth]{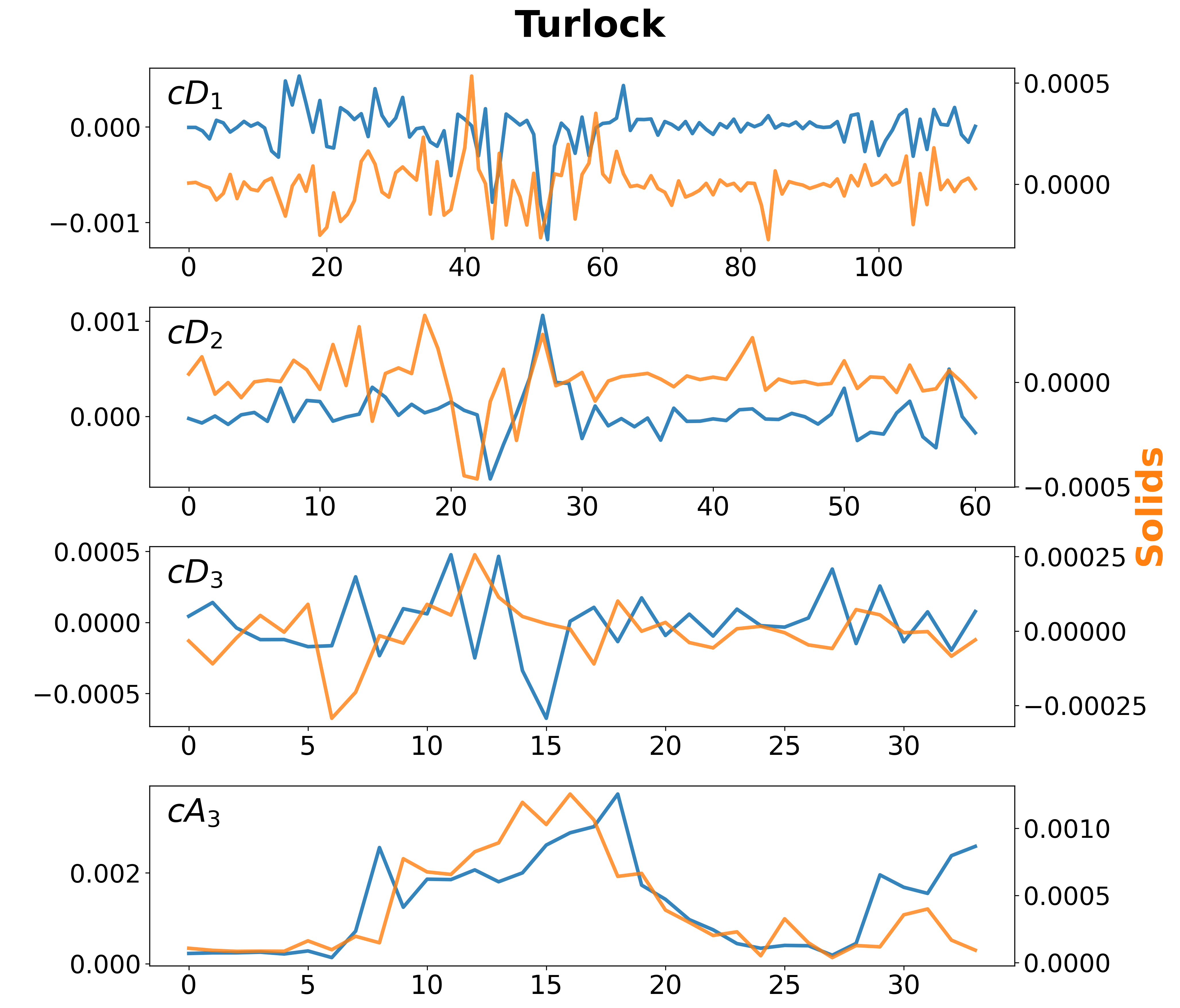}
\includegraphics[width=0.54\linewidth]{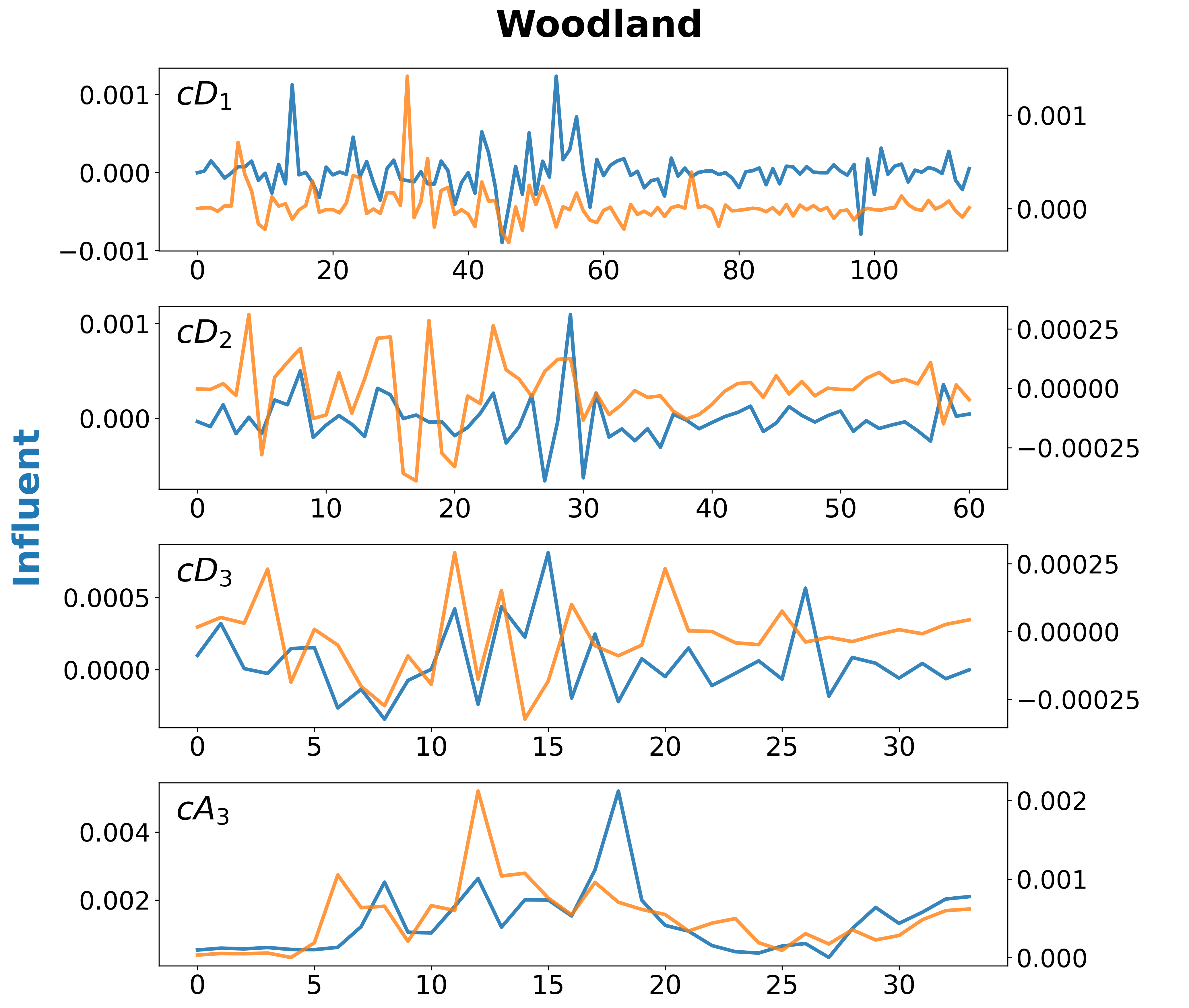}\includegraphics[width=0.54\linewidth]{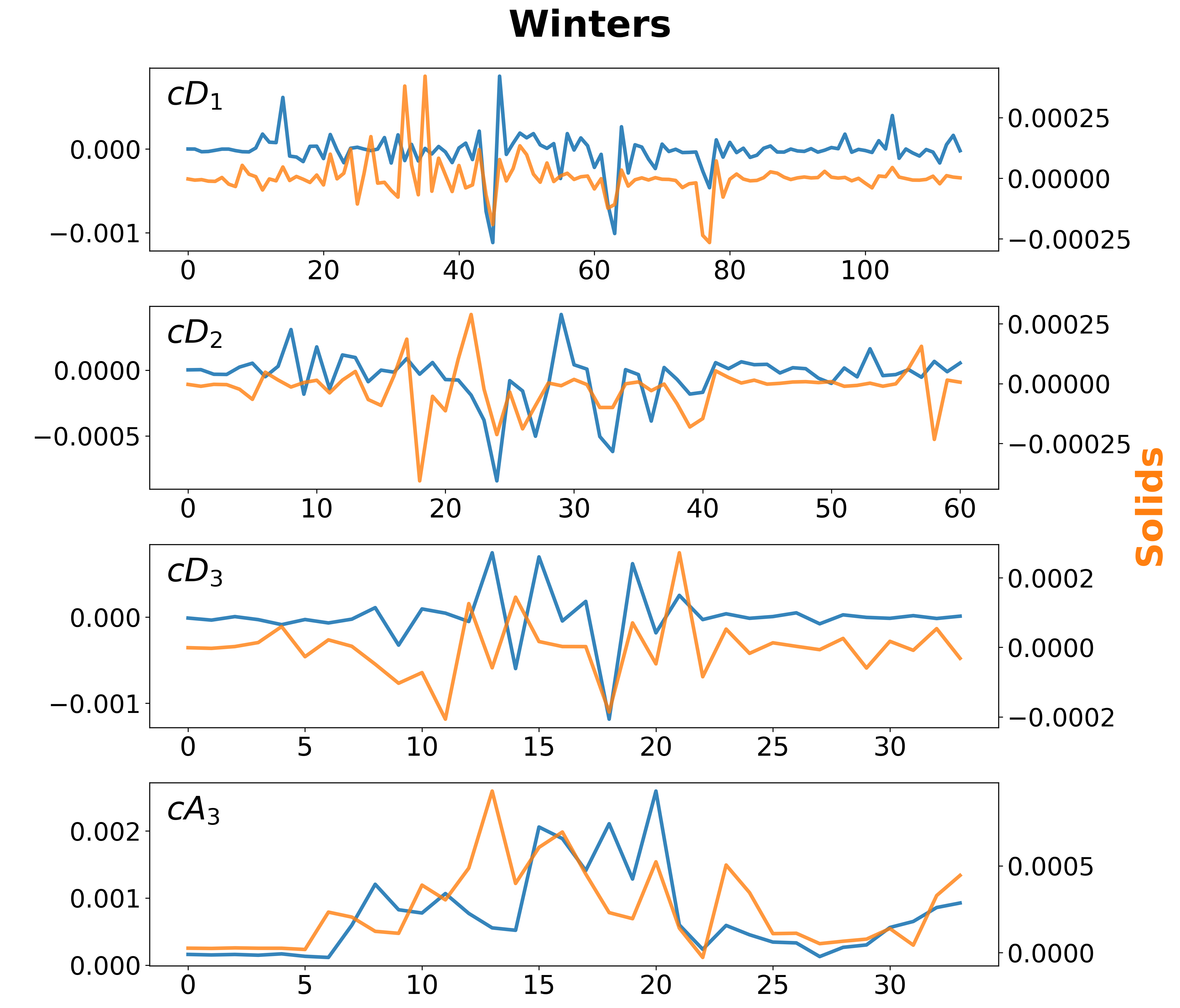}
    \caption{Wavelet decomposition coefficients for SARS-CoV-2 concentration signals from wastewater influent and settled solids samples in Esparto, Turlock, Woodland, and Winters. For each city, the figure displays the detail coefficients ($cD_1$, $cD_2$, $cD_3$) and the approximation coefficient ($cA_3$) obtained from discrete wavelet decomposition. These coefficients represent distinct frequency components of the original time series and are plotted to support the visual comparison of temporal dynamics across sample types and decomposition levels.}
    \label{fig:S_wavelet_coeff_all}
\end{figure}

\begin{figure}[H]
    \centering
    \includegraphics[width=0.53\linewidth]{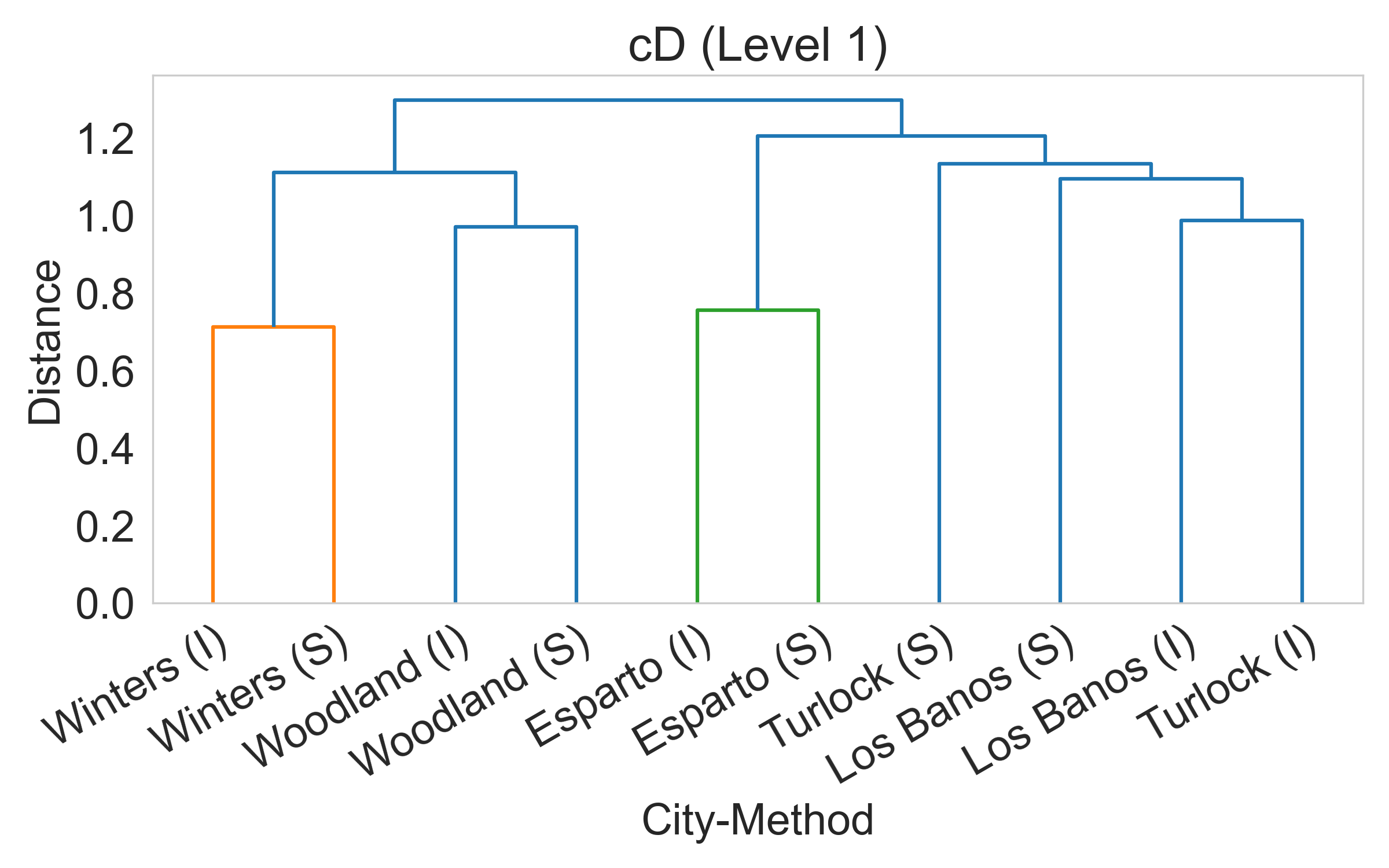}\includegraphics[width=0.53\linewidth]{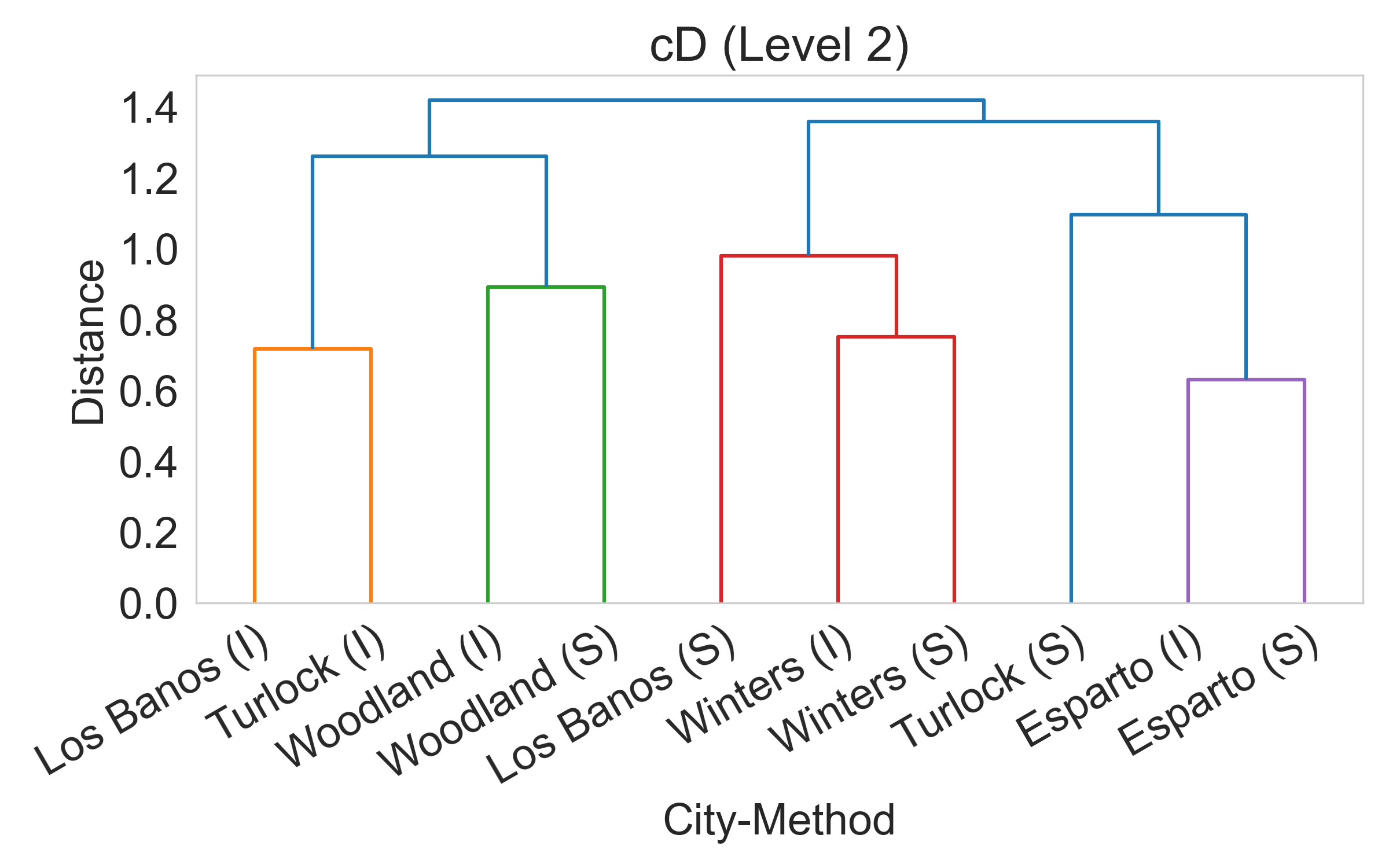}
    \includegraphics[width=0.53\linewidth]{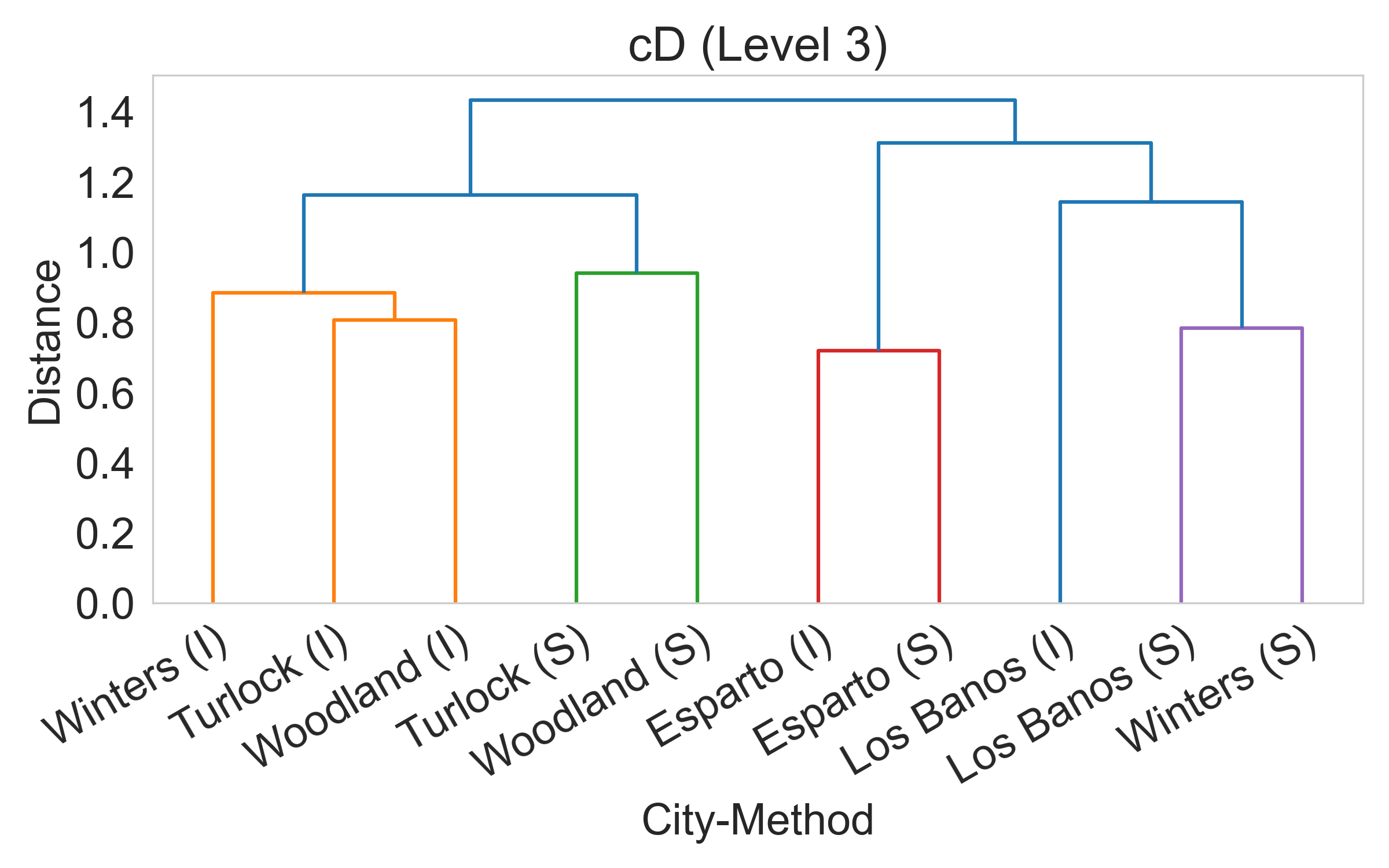}\includegraphics[width=0.53\linewidth]{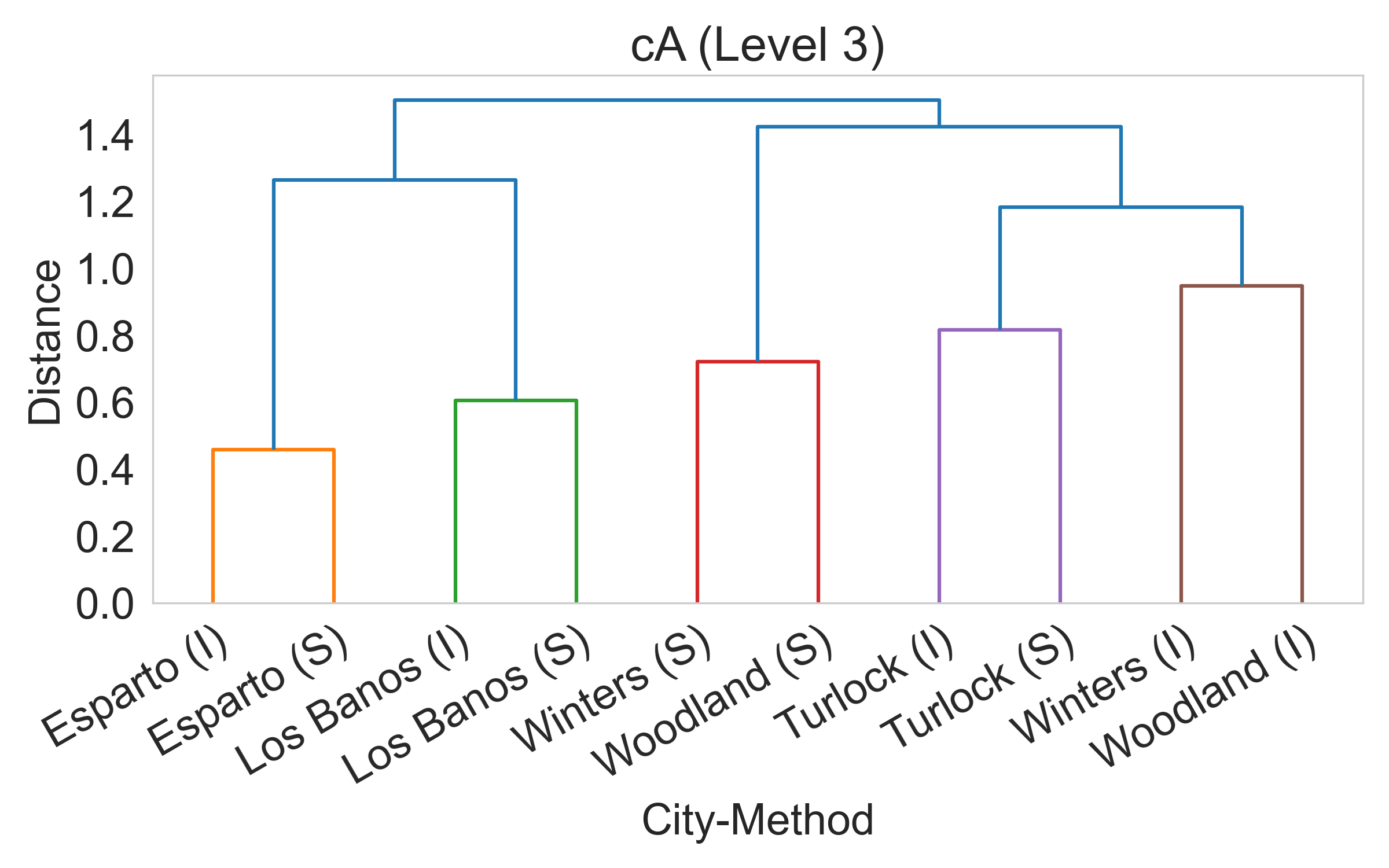}
    \caption{Hierarchical clustering dendrograms of SARS-CoV-2 wastewater signals across cities and sample types using wavelet decomposition coefficients. Clustering results are shown for samples from Los Banos, Turlock, Winters, Woodland, and Esparto, with samples labeled as City(I) for influent and City(S) for solids. Panels represent: {\bf (A)} detail coefficients at level 1 ($cD_1$), {\bf (B)} level 2 ($cD_2$), {\bf (C)} level 3 ($cD_3$), and {\bf (D)} approximation coefficients at level 4 ($cA_4$). Clustering was performed using Ward’s linkage method and correlation distance. Notably, only the $cA_4$ component (Panel D) resulted in consistent clustering by city, grouping influent and solids samples from the same location together—supporting the hypothesis that lower-frequency components capture shared epidemiological signals, while higher-frequency components predominantly reflect noise.}
    \label{fig:S_clusters}
\end{figure}

%%% aknowledge

\end{document}